\documentclass[10pt]{article}

\usepackage{amssymb,amsfonts,amsmath}
\usepackage{graphicx}
\usepackage{subfig}
\usepackage{adjustbox}
\usepackage{colortbl}
\usepackage{xcolor}
\usepackage[left=3cm,right=3cm,top=2cm,bottom=2cm]{geometry}
\usepackage{ifthen}
\usepackage{diagbox}
\usepackage[hidelinks]{hyperref}
\usepackage{wasysym}
\usepackage{authblk}
\newcommand{\ee}{\mathbf{e}}
\newcommand{\yy}{\mathbf{y}}

\newcommand{\Rset}{\mathbb{R}}
\newcommand{\QoI}{\text{QoI}}

\newtheorem{remark}{Remark}

\newcounter{sidenotescounter}

\newcommand{\removecite}[1]{
  \textcolor{red}{$^{***}$}
  \marginpar{
    \begin{flushleft}
      {\scriptsize \textcolor{red}{removed cit} \par}
    \end{flushleft}
  }
}

\DeclareMathOperator*{\argmax}{arg\,max}
\DeclareMathOperator*{\argmin}{arg\,min}

\title{Towards uncertainty quantification of a model for cancer-on-chip experiments}
\author[1]{S. Bertoluzza}
\author[2]{V. Bianchi}
\author[3]{G. Bretti}
\author[1]{L. Tamellini}
\author[4]{P. Zanotti}
\affil[1]{\footnotesize Consiglio Nazionale delle Ricerche -- Istituto di Matematica Applicata e Tecnologie Informatiche, Via Ferrata 5, 27100, Pavia, Italy}
\affil[2]{Università degli Studi di Pavia, Via Ferrata 5, 27100, Pavia, Italy}
\affil[3]{Consiglio Nazionale delle Ricerche -- Istituto di Applicazioni del Calcolo, Via dei Taurini 19, 00185, Rome, Italy}
\affil[4]{Università degli Studi di Milano -- Dipartimento di Matematica, Via Saldini 50, 20133, Milano, Italy}

\date{}

\begin{document}

\maketitle

\begin{abstract}
This study is a first step towards using data-informed differential models to predict and control the dynamics of cancer-on-chip experiments. We consider a conceptualized one-dimensional device, containing a cancer and a population of white blood cells. The interaction between the cancer and the population of cells is modeled by a chemotaxis model inspired by Keller-Segel-type equations, which is solved by a Hybridized Discontinuous Galerkin method. Our goal is using (synthetic) data to tune the parameters of the governing equations and to assess the uncertainty on the predictions of the dynamics due to the residual uncertainty on the parameters remaining after the tuning procedure. To this end, we apply techniques from uncertainty quantification for parametric differential models. We first perform a global sensitivity analysis using both Sobol and Morris indices to assess how parameter uncertainty impacts model predictions, and fix the value of parameters with negligible impact. Subsequently, we conduct an inverse uncertainty quantification analysis by Bayesian techniques to compute a data-informed probability distribution of the remaining model parameters. Finally, we carry out a forward uncertainty quantification analysis to compute the impact of the updated (residual) parametric uncertainties on the quantities of interest of the model. The whole procedure is sped up by using surrogate models, based on sparse-grids, to approximate the mapping of the uncertain parameters to the quantities of interest.
\end{abstract}

\paragraph{Keywords}
Cancer-on-chip, Chemotaxis models (Keller--Segel-type equations), Hybridizable discontinuous Galerkin, Uncertainty quantification, Sparse grids, Bayesian inversion

\paragraph{AMS subject classification}
41A10, % approximation by polynomials
62F15, % Bayesian inference
65M60, %	Finite element, Rayleigh-Ritz and Galerkin methods for initial value and initial-boundary value problems involving PDEs
92-10. % Mathematical modeling or simulation for problems pertaining to biology

% ----------------------------------------------------- %
%                        INTRO
% ----------------------------------------------------- %

\section{Introduction}
\label{sec:introduction}

%\sidenotepietro{La posizione di tabelle e figure è da ottimizzare in base alla versione finale}
Traditional two-dimensional cell cultures can hardly replicate accurately the complex cell–cell interactions found in the human body,
which is why animal models are often preferred.% \removecite{Beissner.EtAl:16}. 
However, due to the significant differences between animal and human physiology,
animal experiments are increasingly acknowledged as an imperfect representation of the human system %\removecite{Chan.EtAl:13}
and are always accompanied by ethical concerns. In recent years, organs-on-chip technologies,
i.e. microfluidic devices engineered to reproduce the behavior of entire living organisms
and their surrounding environment with modular degree of complexity \cite{Beissner.EtAl:16} 
%\removecite{Gori.EtAl:16, Mittal.EtAl:19, Monteduro.EtAl:23, Tsai.Trubelja.Shen.Bao:17} 
have therefore been increasingly employed as experimental tools allowing the reproduction of in-vivo systems on a in-vitro system.

In this work we focus in particular on laboratory experiments conducted on patient-derived human cells in the framework of the Cancer-on-Chip (CoC) technology 
%\removecite{DeNinno.EtAl:21, Maulana.EtAl:21, Vacchelli.EtAl:15}
\cite{Vacchelli.EtAl:15}, 
an immune-oncology microfluidic chip designed to gain insights into the fundamental mechanisms of immunocompetent behavior.
Starting from biological hypotheses and thanks to the availability of data from laboratory experiments (e.g. video footages),
mathematical models able to replicate biological experiments can be developed.
Once validated and calibrated against data, such models can help unveiling the complex, nonlinear and stochastic nature of cell dynamics,
and they can contribute to the development of biological digital twins.
Digital twins can be seen as a sustainable and non-invasive computational tools for
supporting biological experiments, thanks to their low economic impact and
easy reproducibility. They allow better understanding on how perturbations in these complex systems may contribute to diseases and
support the experimental research on drug testing of new and effective treatments.
However, because of the intrinsic complexity and heterogeneity of biological processes,
many aspects still remain not fully understood and a variety of mathematical models can be formulated by different approaches: 
microscopic \cite{Lee.etAl:18}, %\removecite{Checcoli.EtAl:20,Lee.etAl:18,Lee.EtAl:20}
macroscopic \cite{Braun.Bretti.Natalini:21}, %\removecite{Braun.Bretti.Natalini:21,Gamba.EtAl:03,Lewin.EtAl:22}
kinetic models \cite{Othmer.Hillen:02} 
and their hydrodinamic/diffusive limits \cite{Natalini.Paul:22}, %\removecite{Bretti.Gosse:21,Dolak.Schmeiser:05, Natalini.Paul:22}
discrete-in-continuous/hybrid models \cite{Bretti.EtAl:21}, %\removecite{Bretti.EtAl:21,DiCostanzo.Natalini.Preziosi:15}
cellular automata \cite{Pompa.Torre.Bretti.DeGaetano:23}.%\removecite{Bretti.DeGaetano:22,Pompa.Torre.Bretti.DeGaetano:23}

In this work we focus on a macroscopic mathematical model for CoC laboratory experiments.
To our knowledge, \cite{Braun.Bretti.Natalini:21} is the first macroscopic model for CoC,
inspired by the Keller-Segel model \cite{Keller.Segel:70}, and based on reaction-diffusion equations with chemotaxis.
Such model describes cell death processes, effects of chemoattractants, interactions,
and competition between two populations of cells coexisting together,
i.e. cancer cells and immune cells, represented by densities. 
Due to the intricacies of the underlying physiological processes, the equations describing CoC experiments
are unavoidably complex, which poses challenging in their numerical approximation as well as in the calibration needed to obtain the final biological digital twins.

As for the first issue, i.e., the numerical approximation of the solution of the models,
the use of both accurate and efficient numerical algorithms is crucial: in this work,
we resort to an ad-hoc Hybridizable Discontinuous Galerkin (HDG) solver \cite{Bertoluzza.Bretti.Pennacchio.Prudhomme:25+}.

Regarding the second issue, i.e., the calibration of the model, we consider an Uncertainty Quantification (UQ) framework with a Bayesian approach
%\removecite{Ghanem.Higdon.Owhadi:17,Smith:13,Sullivan:15}
\cite{BuiThanh.Ghattas.Martin.Stadler:13,Petra.Martin.Stadler.Ghattas:14}.
%\removecite{BuiThanh.Ghattas.Martin.Stadler:13,Petra.Martin.Stadler.Ghattas:14,Stuart:10}
Instead of computing a single value for the parameters of interest, we compute 
a probability density function that assigns larger probability to values of the parameters that are more consistent with the available data.
Such probability density function can be understood as a residual uncertainty after calibration, that can be propagated to the final prediction of
the model by forward UQ techniques.
Before tackling model calibration, we perform a preliminary global sensitivity analysis \cite{Saltelli:08}
% \removecite{Iooss.Lemaitre:15,Saltelli:08, Saltelli.Tarantola.Campolongo.Ratto:04}
to determine which parameters actually contribute to the variability of the system response that will be compared against the data during calibration,
and are hence amenable to calibration: any attempt at calibrating parameters with negligible impact on the system would be instead clearly unsuccessful.
The sensitivity analysis is conducted by computing both Sobol and Morris indices (as well as by a graphical one-at-a-time analysis),
to make its finding more robust. Our methodology can therefore be summarized in the following three-steps workflow
\begin{equation}
\label{eq:UQ-workflow}
    \text{sensitivity analysis} 
    \qquad \longrightarrow \qquad
    \text{Bayesian inversion}
    \qquad \longrightarrow \qquad
    \text{forward UQ}.
\end{equation}
Each step requires several calls of the HDG solver for different combinations of the uncertain parameters. We speed up the workflow by resorting to suitable surrogates of the HDG approximation of the quantities of interest of the model. Indeed, after offline training, the surrogates still accurately approximate the quantities of interest and their evaluation is much less computationally demanding.
Among the several options available in literature \cite{Ghanem.Higdon.Owhadi:17},
%\removecite{Ghanem.Higdon.Owhadi:17,Gramacy:20,Hesthaven.Rozza.Stamm:15,Quarteroni.Manzoni.Negri:15} 
we build the surrogates by means of sparse-grids, which are well-established for UQ \cite{Babuska.Nobile.Tempone:10}.
%\removecite{Babuska.Nobile.Tempone:10,Gerstner.Griebel:03,Piazzola.Tamellini:24,Xiu.Hesthaven:05} 

Former uses of the workflow \eqref{eq:UQ-workflow} are countless.
% . In particular, we closely follow \removecite{Chiappetta.EtAl:23,Li.EtAl:25}. 
To the best of our knowledge though, this is the first application of the entire workflow to a CoC model. The most related works are:
\begin{itemize}
\item \cite{Braun.Bretti.Natalini:22,Bretti.EtAl:2024,Pompa.Torre.Bretti.DeGaetano:23} performing local
  sensitivity analysis, i.e., studying the effect of small perturbations in the parameter values one at a time
  around a fixed nominal setting;
\item \cite{Campanile.Colombi.Bretti:24} performing a global sensitivity analysis of the Keller--Segel equations;
\item \cite{Both.Choudhury.Sens.Kusters:21,Lee.EtAl:23,Psarellis.EtAl:24} employing deep-learning-based model-discovery
    and gaussian process regression to learn the functional form of the one-dimensional Keller--Segel equations and the associated coefficients;
\item \cite{Collin.Kritter.Poignard.Saut:21,Flavien.BenMansour.Mazen:25} recasting the inverse-forward UQ problem as a joint, sequential state-parameter estimation problem.
\end{itemize}

\paragraph{Contribution and outlook} This study is a proof of concept, i.e., a first step towards using data-informed differential models to predict and control the dynamics of CoC experiments. We test the application of the workflow \eqref{eq:UQ-workflow} for a simplified one-dimensional version of the model proposed in \cite{Braun.Bretti.Natalini:21}. For the Bayesian inversion, we use synthetic data, obtained by running the HDG solver for a given combination of the parameters,
and then corrupted by adding gaussian noise. The results in this paper lay the ground for future work, including a more realistic differential model and two-dimensional geometry, enabling the use of real data from CoC experiments.

\paragraph{Organization} The paper is organized as follows:
Section \ref{sec:framework} shortly summarizes the CoC experiment of interest, the corresponding differential model, the HDG solver and the sparse-grid surrogates.
Section \ref{sec:UQ} is concerned with the UQ problem formulation, whose results are then discussed in Section \ref{sec:numerical-results}.
Finally, we end up with conclusions and future perspectives of our research in Section \ref{sec:conclusions}.

% ----------------------------------------------------- %
%                 MODEL AND APPROXIMATION
% ----------------------------------------------------- %

\section{Differential model and numerical approximation}
\label{sec:framework}
In this section we introduce a one-dimensional model for CoC experiments, which is a simplified version of the one proposed in \cite{Braun.Bretti.Natalini:21, Braun.Bretti.Natalini:22}. Then, we discuss numerical techniques to approximate the solution of the model and/or the quantities of interest thereof.

% ----------------------------------------------------- %
%                     EXPERIMENT
% ----------------------------------------------------- %

\subsection{CoC experiment}\label{sec:coc}

 The development of CoC devices is particularly promising for their ability to replicate the complex cancer microenvironment, where immune cells are a critical component and play a vital role in cancer growth and maintenance. This immunocompetent behavior is of significant interest because immune cells can either attenuate or promote cancer progression \cite{Kumar.Varghese:19}. In this intricate scenario, the migration capabilities of both immune cells and cancer cells are crucial for the inflammatory process. 

The model introduced in \cite{Braun.Bretti.Natalini:21, Braun.Bretti.Natalini:22}, and further simplified in Section \ref{sec:model} below, is inspired by \cite{Vacchelli.EtAl:15}
%the work of Businaro-Vacchelli et al. \removecite{Businaro.EtAl:13,Vacchelli.EtAl:15} 
for its paradigmatic relevance in the study of interactions between cancer cells and immune system cells. In the laboratory experiment, cancer cells are treated with a chemotherapy drug, are dying and quasi-static, and secrete a chemical signal that acts as chemoattractant for wild-type immune cells. More specifically, the immune population consists of peripheral blood mononuclear cells, including different cell species: monocytes, dendritic cells, and T and B lymphocytes.

\begin{figure}[tp]
  \centering
    $\vcenter{\hbox{\includegraphics[scale=0.6]{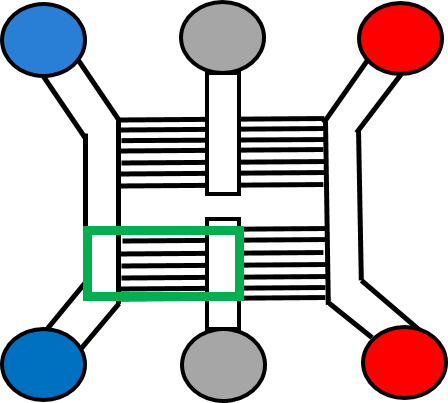}}} $
    \hspace{25pt}
    $\vcenter{\hbox{\includegraphics[scale=0.85]{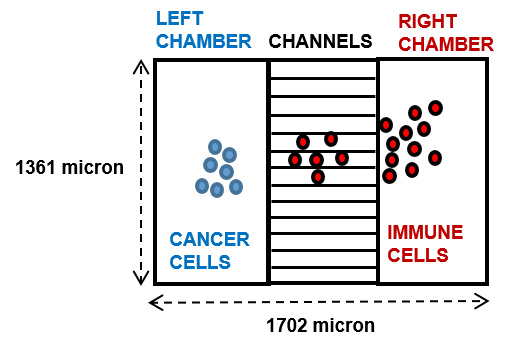}}}$
    \hspace{25pt}
    %$\vcenter{\hbox{\includegraphics[scale=0.45]{images/chip_monitored_area.png}}}$
  \caption{Two-dimensional planimetry of the CoC device (left), with blue and red wells initially filled with cancer and immune cells, respectively. The area delimited in green (left) and zoomed in (right) is the monitored part during the laboratory experiment.}
  \label{fig:expGeometry}
\end{figure}

The microfluidic device of CoC experiment is schematically represented in Figure~\ref{fig:expGeometry}: on the left the whole chip geometry is represented, while on the right the area monitored in the video footage of the laboratory experiment is depicted.
%%%%%
%Immune compartment (left): 1000 μm (width) × 100 μm (height)
%Tumor compartment (right): 1000 μm (width) × 100 μm (height)
%Migration channels:
%Width: 12 μm
%Length: 500 μm
%Height: 10 μm
%Left chamber: Loaded with 2×10⁶ immune cells (murine splenocytes or human PBMCs) in 200 μL medium
%Right chamber: Loaded with 5×10⁴ tumor cells, either untreated or pre-treated with anthracyclines (DOXO or MTX)

More in detail, the CoC device consists of two closed-ended central compartments and two cell culture compartments (see left panel of Figure \ref{fig:expGeometry}), allowing circulation of cells and chemical signals in all interconnected areas of the chip. The cell compartments are connected to the central compartments via four sets of micron-sized channels. Cancer cells are introduced into the two upper and lower reservoirs of the left cell culture compartments, while immune cells, labeled with a fluorescent tracker, are placed in the two reservoirs of the right chamber. The closed-ended central channels, containing medium alone, are designed as buffer chambers to regulate experimental conditions and prevent floating immune cells from flowing directly into the cancer compartment upon initial loading.

The monitored area of the microfluidic device consists of two main culture
chambers (see right panel in Figure~\ref{fig:expGeometry}): the left chamber hosting adherent cancer cells and the right chamber hosting floating immune cells, connected by microchannels allowing chemical and physical contacts. Immune cells, start migrating driven by the chemical signal produced by cancer cells. After crossing microchannels, they reach cancer cells and interact with them.

% ----------------------------------------------------- %
%                    Model
% ----------------------------------------------------- %
\subsection{Differential model}\label{sec:model}

We consider a simplified version of the model introduced in \cite{Braun.Bretti.Natalini:21,Braun.Bretti.Natalini:22} for the dynamics in a CoC device, inspired by the Keller-Segel equations \cite{Keller.Segel:70}. Our model consists of two weakly coupled diffusion-advection-reaction
partial differential equations. The model describes the evolution in space and time of the chemoattractant secreted by the cancer cells and of the immune cells that are
attracted by the chemoattractant. As the cancer cells are treated with chemotherapy drugs, they do not proliferate and the chemoattractant production decreases in time.

We simplify the complex geometry of CoC devices by considering a one-dimensional spatial domain. The dimensional reduction is a deliberate choice for this preliminary study, in order to keep the computational complexity to a minimum, while setting up the UQ framework. Despite of this simplification, the model is still able to capture the key patho-physiological aspects underlying the experiment, and to provide some initial relevant insights into the influence of some parameters on the dynamics. 
In the context of the CoC experiment described in Section~\ref{sec:coc}, we consider as the computational domain a one-dimensional horizontal section of the CoC monitored area, cf. the right panel of Figure~\ref{fig:expGeometry}. We assume that the size of the section is \( L= 1000 \mu\text{m}\) in order to have a computational domain reproducing the length on the horizontal axis of two compartments (i.e. the 
cumulative length of the left chamber and the microchannels) hosting the two different populations.

Denoting by $[0,L]$ the space domain, we assume that cancer cells are concentrated on the left side around the point $x=250$ and produce chemoattractant for $t>0$. The immune cells are on the right, around $x = 750$. This configuration mimics the initial
state of the experiment described above. We consider the time interval $[0, T]$ with $T = 1800s$, which is much shorter than in standard CoC experiments but sufficiently long to observe the relevant dynamics of our simplified model.

Denote by $u$ and $\varphi$ the concentration of the immune cells and of the chemoattractant, measured in moles per unit length. With the above simplifications and assumptions, the model in \cite{Braun.Bretti.Natalini:21,Braun.Bretti.Natalini:22} is reduced to the following initial-boundary value problem.
\begin{equation}
\begin{alignedat}{2}
  \partial_tu - \partial_{x} \big( \nu \partial_xu - \chi u \partial_x\varphi\big) &= f_u \quad & \text{in } (0,L)\times(0,T) &\\ 
  \partial_t\varphi - \partial_{x} \big(\mu\partial_x\varphi \big) + a\varphi &= f_{\varphi} &  \text{in } (0,L) \times (0,T) &\\[4pt]
  u(\cdot,0) &= u_0  & \text{in } (0,L) &  \\
  \varphi(\cdot,0) &= \varphi_0 &  \text{in } (0,L)\\[4pt]
  \partial_xu(0,\cdot) = \partial_xu(L,\cdot) &= 0 &  \text{in } (0,T)\\
  \partial_x\varphi(0,\cdot) = \partial_x\varphi(L,\cdot) &= 0 & \text{in } (0,T)
\label{eq:modello2eq}
\end{alignedat}
\end{equation}

\begin{figure}[tp]
  \centering
    \includegraphics[scale=0.45]{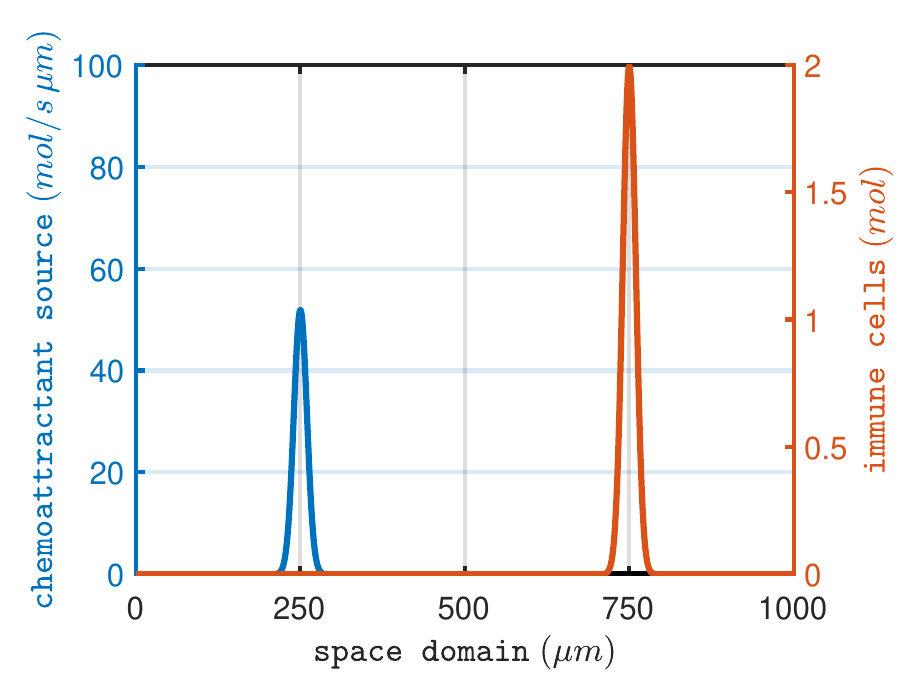} 
    \caption{Schematic representation of the one-dimensional space domain with the initial profile of the chemoattractant source and of the immune cells distribution for a reference parameter combination.}
\label{fig:expapprox}
\end{figure}

In contrast to the original Keller-Segel model, the partial differential equations for $u$ and $\varphi$ are only one-way coupled. The coupling is realized via the chemotactic (advection) term $ \partial_x(\chi u \partial_x\varphi)$, which accounts for the migration of the immune cells towards regions with higher chemoattractant concentration. The constant $\chi$ denotes the \emph{chemotactic sensitivity coefficient}. The second-order terms $\partial_x(\nu \partial_xu)$ and $\partial_x(\mu \partial_x\varphi)$ describe diffusion according to Fick's law, whereas $a\varphi$ represents the natural degradation of the chemoattractant. We set the two forcing terms as follows:
\begin{equation}
        f_u(x,t) = 0 \qquad \text{and} \qquad  \ f_{\varphi}(x,t) = \frac{k_{\varphi}\exp(-\rho t)}{\sqrt{2\pi} \sigma_\varphi}   \exp\left(-\frac{(x - \ c_{\varphi})^2}{2\sigma_{\varphi}^2}\right).
    \label{eq:source}
\end{equation}
The first one is null as there is no immune cells source. The second one accounts for the chemoattractant production by the cancer cells.
The Gaussian dependence on the space variable models the position of the cancer cells, see Figure \ref{fig:expapprox},
which are assumed to remain still. The exponential decay in time represents the weakening of the cancer cells due to drug administration. 
The initial data are defined as
\begin{equation}\label{eq:u0_phi0}    
        u_0(x) = \frac{k_{u_0}}{\sqrt{2\pi} \sigma_{u_0}} \exp\left(-\frac{(x-c_{u_0})^2}{2 \sigma_{u_0}^2}\right)
        \qquad \text{and} \qquad\ \ 
        \varphi_0(x) = 0.      
\end{equation}
The first one models the initial distribution of the immune cells by a Gaussian profile, see again Figure~\ref{fig:expapprox}.
The second one indicates the absence of chemoattractant at the initial time. Finally, homogeneous Neumann boundary conditions impose that the system is closed.

Table~\ref{tab:parametri} summarizes the parameters involved in the equations \eqref{eq:modello2eq}-\eqref{eq:u0_phi0}, and highlights the respective roles and physical units. When available, specific parameter values or ranges of variability are taken from the literature. For the other parameters, related to specific properties of the model proposed in this section, the value or the range of variability are determined by design choice (DC), i.e. based on our previous experience with the dynamics of the model. 

\begin{table}[tp]
\centering 
\begin{adjustbox}{width=0.95\textwidth} 
    \begin{tabular}{cccccc} 
    \hline 
    \textbf{Param.} & \textbf{Description} 				& \textbf{Range} 	& \textbf{Value} 	& \textbf{Unit} & \textbf{Ref} \\ 
    \hline 
    $\mu$ 	& Diffusivity of the chemoattractant 			& $[700 - 1100]$	& - 			& $\mu m^2/\mathrm{s}$ & \cite{Murray:03} \\ % value = $900$ 
    %\hline 
    $\nu$ 	& Diffusivity of immune cells 				& $[100 - 300]$		& - 			& $\mu m^2/\mathrm{s}$ & \cite{Murray:03} \\ % value = $200$
    %\hline 
    \(a\) 	& Consumption rate of \(\varphi\) 			& - 			& \(10^{-4}\) 		& \(1/s\) & \cite{Curk.Marenduzzo.Dobnikar:13} \\ 
      $\chi$ 	& Chemotactic sensitivity coefficient 			& - 			& $5\times 10^{-2}$ 	& $\mu m^3 /(s\,mol)$ 		& (DC) \\ 
    %  \hline 
      $k_{u_0}$ 	& Amplification coefficient of \(u_0\) 			& - 			& $50$ 			& $mol/s$ 		& (DC) \\ 
    %  \hline 
      $c_{u_0}$ 	& Center of the initial condition for immune cells 	& - 			& $750$		 	& $\mu \mathrm{m}$ 	& (DC) \\ 
    %  \hline 
      $\sigma_{u_0}$ & Spread of the initial condition for immune cells 	& - 			& $10$ 			& $\mu \mathrm{m}$ 	& (DC) \\ 
    %  \hline 
      $k_{\varphi}$ & Amplification coefficient of \(f_{\varphi}\) 	& $[600 - 2000]$ 		&  -			& $mol/s$ 		& (DC) \\  % value = 50
    %  \hline 
      $\rho$ 	& Decay rate of dying cancer cells caused by drug & $[10-40]\times 10^{-4}$ & -  			& $1/\mathrm{s}$ 	& (DC) \\  % value = 27.5
    %  \hline 
      $c_{\varphi}$ & Center of the chemoattractant source term \(f_{\varphi}\) & $[150 - 350]$ 	& - 		 	& $\mu \mathrm{m}$ 	& (DC) \\% value = ${3\over4}L = 750$
    %  \hline 
      $\sigma_{\varphi}$ & Spread of the chemoattractant source term $f_{\varphi}$ & $[5 - 15]$& 		- 	& $\mu \mathrm{m}$ 	& (DC) \\ % val=10
      \hline 
    \end{tabular}
\end{adjustbox}
\caption{Summary of the parameters involved in \eqref{eq:modello2eq}-\eqref{eq:u0_phi0}.} 
\label{tab:parametri} 
\end{table} 

In the subsequent UQ analysis, we are mostly interested in two Quantities of Interest (QoI), i.e. in two functionals of the solution $(u, \varphi)$ of \eqref{eq:modello2eq}. First, $M:[0,T] \to \mathbb{R}$ is the center of mass (in space) of the immune cells over time and is defined as
\begin{equation}\label{eq:center-of-mass}
M(t) := \frac{1}{U} \int_{0}^{L} xu(x,t) \, dx, \quad \text{where} \quad
U := \int_0^L u(x,t) \, dx = \int_0^L u_0(x) \, dx.
\end{equation}
where the second equality holds true due to \eqref{eq:source} and the boundary conditions.
This is our paradigm of a QoI that can be inferred from experimental data, e.g. from tracking analysis of time-lapse video recordings.
The second QoI is the total amount (in space) of chemoattractant $I:[0,T] \to \mathbb{R}$ over time, obtained as
\begin{equation}
  \label{eq:integral}
    I(t) := \int_0^{L} \varphi(x,t) \, dx.
  \end{equation}
This exemplifies a quantity that is useful to know but that can hardly be inferred by direct experimental measurements. Actually, one could be interested in the amount of chamoattraftant only in a subset of the space domain. We work on the entire domain so that an explicit formula for $I$ can be established and used to test the accuracy of the HDG solver and of the surrogate models, cf.~\eqref{eq:I-formula} below.

% ----------------------------------------------------- %
%                        HDG
% ----------------------------------------------------- %

\subsection{HDG solver}
\label{sec:HDG}

We approximate the solution of the initial-boundary value problem \eqref{eq:modello2eq} by a MATLAB implementation of the HDG solver developed in \cite{Bertoluzza.Bretti.Pennacchio.Prudhomme:25+}. We refer to \cite{Du.Sayas:19} for a general introduction to HDG solvers. The following properties motivate the use of a HDG solver for this problem. First, the approximate solution satisfies counterparts of the balance laws underlying the first two equations in \eqref{eq:modello2eq} at the element level. In particular, the total amount $U$ of immune cells is constant in time, see \eqref{eq:center-of-mass}, and its approximation is guaranteed to remain constant in time as well. Second, stabilization techniques are well-established for the notoriously critical convection-dominated regime, i.e. when the dynamic is mostly driven by the transport term $\partial_x(\chi u \partial_x\varphi)$ in the first equation of \eqref{eq:modello2eq}. Finally, the HDG paradigm allows for a simple imposition of interface conditions among different subdomains. This property is not relevant for the one-dimensional model in Section~\ref{sec:model}, but it is of paramount importance for future extension to the geometry of a realistic a CoC device, where one-dimensional (microchannels) and two-dimensional (compartments) components must be coupled, cf. Figure~\ref{fig:expGeometry}. In an HDG discretization framework, such kind of coupling can be easily carried out using the approach proposed in \cite{Bertoluzza.EtAl:23}.

The HDG solver results from the discretization of the weak formulation obtained by introducing in \eqref{eq:modello2eq} the fluxes $j$ and $\psi$ of immune cell and chemoattractant density, respectively. Note that the partial differential equations in \eqref{eq:modello2eq} are rewritten in terms of fluxes as
\begin{equation}
\begin{alignedat}{3}
  \partial_t u + \partial_{x} j 
  &= 0
  & \quad \text{and} \quad  
  & \frac{1}{\nu} j + \partial_x u + \frac{\chi}{\nu} u \partial_x \varphi \;\,
  &= 0 \quad 
  & \text{in } (0,L)\times(0,T), \\ 
  \partial_t \varphi + \partial_{x} \psi + a\varphi 
  &= f_\varphi
  & \quad \text{and} \quad 
  &\frac{1}{\mu} \psi + \partial_x \varphi  
  &= 0 \quad 
  & \text{in } (0,L) \times (0,T).
\label{eq:modello2eq-fluxes}
\end{alignedat}
\end{equation}
Analogously, the boundary conditions become
\begin{equation}
j(0, \cdot) = j(L, \cdot) = \psi(0, \cdot) = \psi(L,\cdot) = 0 \quad \text{in } (0, T).
\label{eq:modello2eq-fluxes-BCs}
\end{equation}

\newcommand{\hu}{\widehat u}
\newcommand{\fu}{j}
\newcommand{\p}{\varphi}
\newcommand{\fp}{\psi}
\newcommand{\Gridh}{{\mathcal{T}_h}}
\newcommand{\xo}{{x_{k-1}}}
\newcommand{\xu}{{x_{k}}}
\newcommand{\dx}{\partial_x}
\newcommand{\hp}{\widehat\varphi}
\newcommand{\gRu}{??}
\newcommand{\gRp}{??}
\newcommand{\inu}{\frac 1 \nu}
\newcommand{\imu}{\frac 1 \mu}
\newcommand{\Poly}[1]{\mathbb{P}_{#1}}
\newcommand{\dt}{\partial_t}

For the discretization in space, we use a partition $0 = x_0 < x_1\dots < x_{N_\mathtt{s}} = L$ of the space domain, consisting of $N_\mathtt{s}+1$ nodes. We assume the partition to be uniform, i.e.
\begin{equation*}
\label{eq:HDG-meshsize}
x_{k} - x_{k-1} = \frac{1}{N_\mathtt{s}} =: h
\end{equation*}
for $k=1,\dots, N_{\mathtt s} +1$. We associate with each node $x_k$, $k=0,\dots, N_\mathtt{s}+1$, approximations
\begin{equation}
\label{eq:HDG-multipliers}
\hu^k_h, \;\hp^k_h: [0, T] \to \mathbb{R}
\end{equation}
of the point values $u(x_k)$ and $\varphi(x_k)$. Such unknowns play the role of Lagrange multipliers. We associate with each interval $I_k := (x_{k-1}, x_k)$, $k=1,\dots, N_\mathtt{s}+1$, approximations
\begin{equation}
\label{eq:HDG-concentrations-fluxes}
u_h^k,\; \varphi_h^k,\;\psi_h^k: [0,T] \to \mathbb{P}_1(I_k)
\qquad \text{and} \qquad j_h^k: [0,T] \to \mathbb{P}_0(I_k)
\end{equation}
of the concentrations $u$, $\varphi$ and of the fluxes $j$, $\psi$. Here, $\mathbb{P}_\ell(I_k)$, $\ell \in \mathbb{N}_0$, denotes the space of polynomials of degree at most $\ell$ on $I_k$. We introduce also numerical fluxes of $j$ and $\psi$ at the nodes
\begin{equation}
\begin{alignedat}{3}
\label{eq:HDG-numerical-fluxes}
  \widehat \fu^{k}_{+} := - j^{k+1}_h(x_{k}) + \frac{\tau_u} h ( u^{k+1}_h(x_{k}) - \hu^{k}_h),
  & \quad \text{and} \quad  
  & \widehat \fu^k_{-} :=   j^{k}_h(x_k) + \frac{\tau_u} h ( u^{k }_h(x_k) - \hu^k_h),\\ 
  \widehat \fp^{k}_+ := -\fp^{k+1}_h(x_{k}) + \tau_\varphi ( \p^{k+1}_h(x_{k}) - \hp^{k}_h),  
  & \quad \text{and} \quad 
  &\widehat \fp_-^{k} := \fp^k_h(x_k) + \tau_\varphi ( \p^k_h(x_k) - \hp^k_h),
\end{alignedat}
\end{equation}
where $\tau_u, \tau_\varphi > 0$ are penalization parameters. For $k = 0$ (resp. $k=L$) we leverage the Neumann boundary conditions and override the above expressione by setting 
\begin{equation}\label{eq:Neumann}
\fu^{0}_- = \fp^0_- = 0, \qquad \text{(resp. $\fu^{N_\mathtt{s}}_+ = \fp^{N_\mathtt{s}}_+ = 0$).}
 \end{equation}

The unknowns in \eqref{eq:HDG-multipliers}-\eqref{eq:HDG-concentrations-fluxes} are obtained via the following semi-discretization of \eqref{eq:modello2eq-fluxes}
\begin{equation}
\label{eq:HDG-modello2eq-fluxes}
\begin{alignedat}{3}
    \forall w \in \mathbb{P}_1(I_k) &\qquad&
    \int_{I_k} \Big(\partial_t u_h^k \,w - \fu^k_h  \,\dx w\Big )  + \widehat \fu^{k}_- \,w(x_k) + \widehat \fu^{k-1}_+ \,w(x_{k-1})  
    &= 0,\\
    \forall w \in \mathbb{P}_1(I_k) &\qquad&
    \int_{I_k}  \Big(\partial_t \p^k_h\, w - \fp^k_h \,\dx w + a \varphi^k_h \,w\Big)+ \widehat \fp^k_- \,w(x_k) + \widehat \fp^{k-1}_+ \,w(x_{k-1})    
    &=  \int_{I_k} f_\p  w,\\
    \forall q \in \mathbb{P}_0(I_k) &\qquad&
    \int_{I_k} \Big(\inu\fu^k_h \,q + \frac \chi \nu \imu \fp^k_h \,u^k_h \,q\Big)+ \hu^k_h\,q(x_k) - \hu^{k-1}_h\,q(x_{k-1}) 
    &= 0,\\
    \forall w \in \mathbb{P}_1(I_k) &\qquad&
    \int_{I_k} \Big(\imu \fp^k_h\, w -  \p^k_h \, \dx w \Big) +  \hp^{k}_hw(x_k) -  \hp^{k-1}_h w(\xo) 
    &= 0,
\end{alignedat}
\end{equation}
for $k=1,\dots, N_\mathtt{s}$, and the following semi-discretization of \eqref{eq:modello2eq-fluxes-BCs}
\begin{equation}
\label{eq:HDG-modello2eq-fluxes-BCs}
    \widehat \fu^k_-  + \widehat \fu^{k}_+  = 0, \qquad\qquad
\widehat \fp^k_- + \widehat \fp^k_+ = 0, 
\end{equation}
for $k=0,\dots, N_\mathtt{s}$. Remark that, in view of   \eqref{eq:Neumann}, for $k = 0$ and $k = N_\mathtt{s}$, the above expression encode the homogeneous Neumann conditions.

We resort to a full discretization by applying the backward Euler scheme with $N_\mathtt{t}$ uniform time steps in \eqref{eq:HDG-modello2eq-fluxes}-\eqref{eq:HDG-modello2eq-fluxes-BCs}. The approximations at the initial time are obtained by $L^2$-orthogonal projections and point evaluations of the initial values \eqref{eq:u0_phi0}. For the approximations at the successive time steps, we solve the nonlinear system of algebraic equations resulting from \eqref{eq:HDG-modello2eq-fluxes}-\eqref{eq:HDG-modello2eq-fluxes-BCs}, after static condensation, by means of Newton's iteration, combined with the MATLAB default (backslash) linear solver.

% ----------------------------------------------------- %
%                   SPARSE GRIDS
% ----------------------------------------------------- %

\subsection{Sparse grids surrogates}
\label{sec:surrogates}
The UQ workflow \eqref{eq:UQ-workflow} builds upon the evaluation of the QoIs $M$ and $I$ from \eqref{eq:center-of-mass} and \eqref{eq:integral} for different combinations of the uncertain parameters listed in Table~\ref{tab:parametri}, see Section~\ref{sec:UQ} for details. As the number of such combinations is quite high, it is convenient approximating the QoIs by surrogates rather than by running the HDG solver for each combination.
Indeed, the surrogates can accurately approximate the QoIs
at a small fraction of the cost for running the HDG solver: this speed-up comes at the price of preliminarily building the surrogates by running the solver for a carefully selected set of parameter combinations (\emph{offline training phase}). More precisely, we build surrogates of the QoIs evaluated at each time step, i.e. $M(t_n)$ and $I(t_n)$ for $n=0,\dots, N_\mathtt{t}$.  

\begin{figure}[tp]
  \centering
  \includegraphics[width=0.25\linewidth]{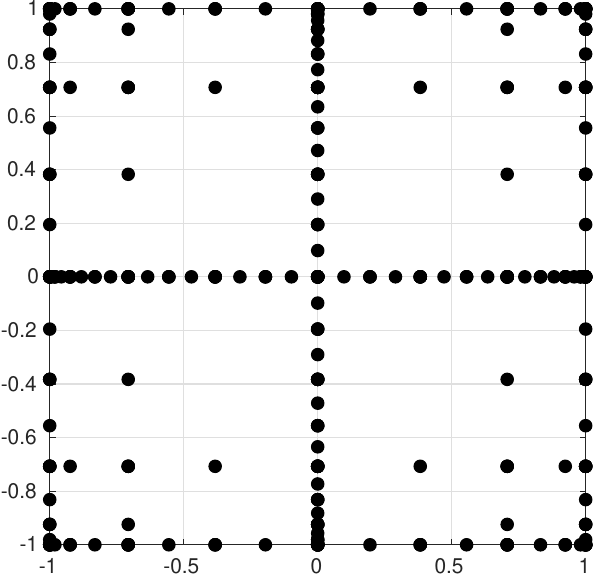}\qquad
  \includegraphics[width=0.25\linewidth]{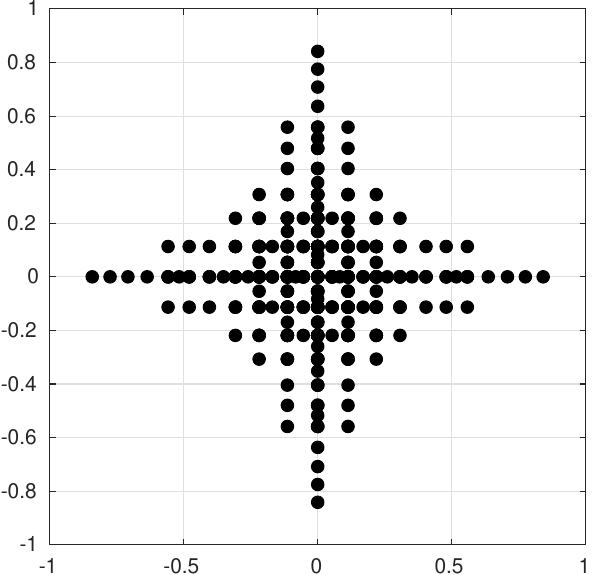}
  \includegraphics[width=0.4\linewidth]{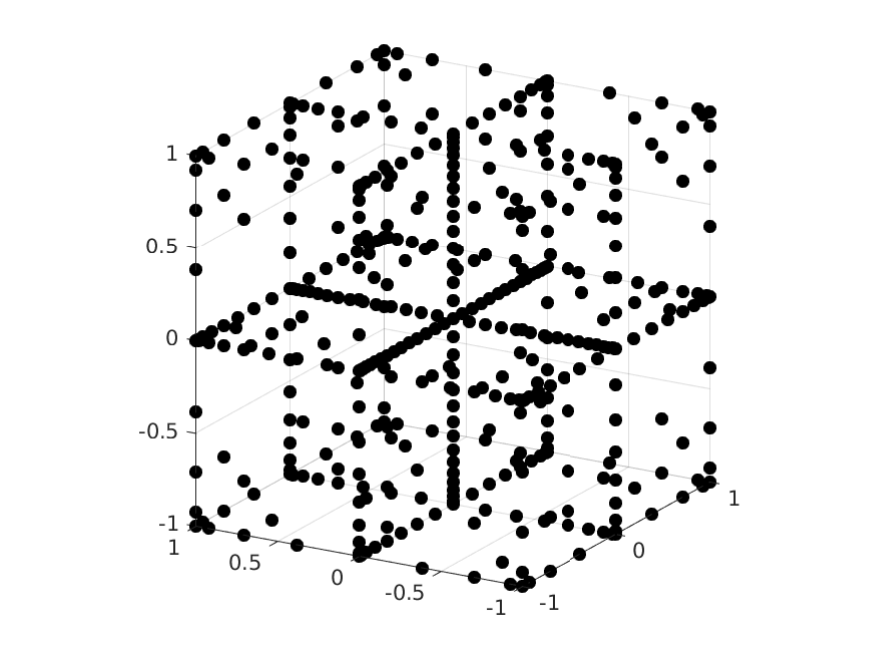}
  \caption{Plot of sparse grids on $[-1,1]^N$ for $w=5$.
    Left: $N=2$, parameters with mutually independent uniform pdf.
    Center: $N=2$, parameters with mutually independent normal pdf with mean zero and standard deviation $\sigma=0.08$.
    Right: $N=3$, parameters with mutually independent uniform pdf.}
  \label{fig:sparse_grids}
\end{figure}

The literature offers many options to build surrogates, see e.g. the review book \cite{Ghanem.Higdon.Owhadi:17}.
%\removecite{Ghanem.Higdon.Owhadi:17,Gramacy:20,Hesthaven.Rozza.Stamm:15,Quarteroni.Manzoni.Negri:15} 
We use the well-established sparse-grid surrogate modeling technique \cite{Babuska.Nobile.Tempone:10},
%\removecite{Babuska.Nobile.Tempone:10,Gerstner.Griebel:03,Xiu.Hesthaven:05}
with the implementation facilities provided by the Sparse Grids Matlab Kit \cite{Piazzola.Tamellini:24}.
This technique builds the surrogates by interpolating the output of the HDG solver for the parameter combinations
(i.e., interpolation points in the parametric domain, see Section \ref{sec:UQ})
chosen according to a specific pattern which, in turn, depends on the probability density function (pdf)
of the uncertain parameters, see e.g. Figure \ref{fig:sparse_grids}.
Note that even for uniform pdf, sparse grids nodes are not equispaced, to prevent from Runge phenomena in the interpolation.
We further elaborate on the selection of the pdf for the parameters in next sections. 

The number of points in the parametric domain to be considered for the interpolation, hence the accuracy of the surrogates,
is governed by a discrete hyperparameter $w \in \mathbb{N}$, called \emph{sparse-grid level}, see \cite[Eq. (15)--(20)]{Piazzola.Tamellini:24}. 
We report in Table~\ref{tab:w_sparsegrids} the number of interpolation points that are required for the values of $w$ and of the number $N$ of parameters that are considered in our subsequent discussion. Note that each run of the HDG solver for a specific interpolation point is completely independent of the other ones. Therefore, the construction of the surrogates can be parallelized. 

\begin{table}[tp] 
\centering 
\begin{adjustbox}{width=0.5 \textwidth}
  \begin{tabular}{c|ccccc} 
    \hline 
    \diagbox[width=5em]{$N$}{$w$} & 1 	& 2 	& 3 	& 4 	& 5 \\
    \hline
    $4$ 	& 9	&  41  	& 137	& 401	& 1105  \\
    $6$ 	& 13  	&  85   & 389   & 1457  & 4865 \\
    \hline 
  \end{tabular}
\end{adjustbox}
  \caption{Number of interpolation points to build a sparse grid as a function of $w$ and $N$.}
\label{tab:w_sparsegrids}
\end{table}

\section{UQ workflow}
\label{sec:UQ}
In this section we discuss the UQ workflow \eqref{eq:UQ-workflow} in detail, with special emphasis on the goal of each step and the tools to be used therein.

% ----------------------------------------------------- %
%                    FRAMEWORK
% ----------------------------------------------------- %

\subsection{General framework}
\label{sec:UQ-framework}
Recall that the parameters involved in the equations \eqref{eq:modello2eq}-\eqref{eq:u0_phi0} are summarized in Table~\ref{tab:parametri}.
For some of these parameters, considering a range of variability instead of a specific value is a necessary approach dictated by to lack of knowledge;
for others, it is a choice motivated e.g. by design purposes.
We collect all such uncertain parameters into a vector of dimension $N = 6$
\[
\yy = \big(\mu,\,\nu,\,k_{\varphi},\,\rho,\,c_{\varphi},\sigma_{\varphi}\big) \in \Gamma =: \prod_{n=1}^N \Gamma_n \subset \Rset^N
\]
where $\Gamma_n := [a_n,b_n]$, $n=1,\dots, N$, are the ranges reported in Table~\ref{tab:parametri}. 

Crucially, we can regard the QoIs in \eqref{eq:center-of-mass}-\eqref{eq:integral} as functions depending not only on time but also on $\yy$, i.e. we have mappings
\[
\Gamma \times [0,T] \ni (\yy, t) \mapsto M(\yy, t)
\qquad \text{and} \qquad
\Gamma \times [0,T] \ni (\yy, t) \mapsto I(\yy, t).
\]
Then, our goal is to apply the well-established UQ framework (\ref{eq:UQ-workflow})
%\removecite{Ghanem.Higdon.Owhadi:17,Smith:13,Sullivan:15}
to quantify the variability of the total amount of chemoattractant $I$ from \eqref{eq:integral} due to the uncertainty on $\yy$,
upon having tuned the parameters $\yy$ with  data regarding the center of mass $M$ of the immune cells from \eqref{eq:center-of-mass}.
This process exemplifies the quantification of the variability for a QoI that can hardly be measured during experiments, by means of data relative to another QoI.

We begin by assuming that $\yy$ is a random vector with known pdf,
that we call \emph{prior pdf} for reasons that will be clearer in a moment, and that we denote as $\rho_{\text{prior}}$.
We further assume that the $N$ components of $\yy$ are statistically independent. Hence, we have
\begin{equation}
\label{eq:prior-pdf}
\rho_{\text{prior}}(\yy) = \prod_{n=1}^N \rho_{\text{prior},n}(y_n)
\qquad \text{with} \qquad 
\rho_{\text{prior}, n}(y_n) = \frac{1}{b_n-a_n}
\end{equation}
for $\yy = (y_1,\dots, y_N) \in \Gamma$. The expression of the prior marginal pdfs $\rho_{\text{prior},n}$, $n=1,\dots, N$, is justified by assuming that each component of $\yy$ is uniformly distributed in the respective range. In other words, we assume no a priori preference on the possible values that the uncertain parameters can take.

Since the model in Section~\ref{sec:model} is highly conceptualized, we use synthetic data instead of experimental ones to tune the parameters. To generate synthetic data, we consider $K \geq 1$ different time instants $0<t_1 < \ldots < t_K \leq T$ and choose a specific admissible parameter combination $\yy_{\text{true}} \in \Gamma$. We run the HDG solver for $\yy = \yy_{\text{true}}$, we evaluate the QoI $M$ at the prescribed time instants and we perturb the results by random noise. In order to mimic measurement error, we assume that the noise associated with each time instant is a random variable with Gaussian distribution, zero mean and standard deviation $\sigma_\varepsilon > 0$. Moreover, the noises associated with different time instants are statistically independent. Denoting by $M_{\textbf{HDG}}$ the HDG approximation of $M$, the synthetic data are obtained as
\begin{equation}\label{eq:data_model}
M^*_k := M_{\text{HDG}}(\yy_{\text{true}},t_k) + \varepsilon_k
\qquad \text{with} \qquad
\varepsilon_k \sim \mathcal{N}(0,\sigma_\varepsilon^2),  \quad k=1,\dots, K.
\end{equation}

\begin{remark}
In a more realistic scenario with experimental data, a similar setup can be considered, but more sophisticated models are necessary for the errors $\varepsilon_k$. At the very least, they should have nonzero mean, to account for modeling and discretization errors; moreover, it is reasonable to assume that these errors are not constant throughout $\Gamma$, such that the standard deviation $\sigma_{\varepsilon}$ might also need to be enlarged to accommodate for such variability in addition to the measurement error.  
\end{remark}

With this preparation, we are in position to describe the role and the relevance of each step in the UQ workflow \eqref{eq:UQ-workflow}. We begin here with a general discussion and postpone to the next three subsections a detailed presentation of the technical tools to be used in each step.

\begin{itemize}
\item \underline{Sensitivity analysis}. We preliminarily verify if $M$ responds to each uncertain parameter. In case $M$ is insensitive to some parameters, we fix them to a reference value, because it would be hardly possible calibrating them, i.e. reducing their variance by comparing simulations with data. Subsequently, we check if $I$ is sensitive to any of the parameters to which $M$ is insensitive. This would indeed lower our ability to reduce the uncertainty on $I$ based on the available data (in the worst-case scenario in which $I$ depends uniquely on parameters to which $M$ is insensitive the forecoming calibration step would be essentially useless).

    \item \underline{Bayesian inversion}. We define and sample an updated data-informed pdf for the (remaining) uncertain parameters,
      hereby called \emph{posterior pdf} (as opposed to the \emph{prior pdf} introduced earlier, which is ``data-unaware'') and denoted by $\rho_{\text{post}}$.
      The posterior pdf is designed to assign higher probability to the parameter combinations $\yy$ such that 
      \[
        \big\{ M(\yy, t_1),\dots, M(\yy, t_K) \big\} \text{ is ``closer'' to } \big\{ M^*_1,\dots, M^*_K \big\}.
      \]
      This step is more sophisticated than classical parameter calibration
      because in addition to a minimization process that individuates the
      most ``likely'' values, we also provide a full sample of pdf. 
      
    \item \underline{Forward UQ}. For $t \in (0,T]$, we evaluate $I(\cdot,t)$ on the sample of the posterior pdf generated at the previous step.
      We use such evaluations to estimate the pdf $\pi_I(\cdot,t)$ of $I$. In other words,
      we \emph{propagate the uncertainty} from the posterior $\rho_{\text{post}}$ to $\pi(\cdot, t)$.

\end{itemize}

% ----------------------------------------------------- %
%                    SENSITIVITY
% ----------------------------------------------------- %

\subsection{Sensitivity analysis}
\label{sec:sensitivity}

We first assess the sensitivity of our QoIs qualitatively by observing the response curves, i.e. the
plots of the two QoIs with respect to the parameters, when only one parameter changes and the other ones are fixed to the midpoint of the respective range,
Table~\ref{tab:parametri} (one-at-a-time sensitivity analysis).
We then move to two different and well-established quantitative approaches, i.e., computing both Sobol and Morris indices \cite{Saltelli:08}.
%\removecite{Iooss.Lemaitre:15,Saltelli:08, Saltelli.Tarantola.Campolongo.Ratto:04}
Since the two approaches are quite different in spirit, obtaining analogous results would provide clear evidence for marking a parameter as `relevant' or `irrelevant' and for fixing it to a constant value in the latter case. We shortly recall below the definition of the two indices for a generic QoI hereby denoted as
$\yy \mapsto \QoI(\yy)$.
Furthermore, we comment on the comparison of the two approaches.

\paragraph{Sobol indices.} %This is a variance-based technique, similar to an ANOVA decomposition \removecite{archer.saltelli.sobol:anova}.
Sobol indices are typically referred to as global sensitivity indices, because they attempt at quantifying the sensitivity of $\QoI(\yy)$ to $\yy$
across the entire parameter space $\Gamma$.
The variance $\mathbb{V}[\QoI]$ of $\QoI(\yy)$ is taken as the reference value for the overall variability of  $\QoI$, and it is decomposed into the contributions $V_J$ relative to the interactions among any subset of parameters, identified by an index set $J \subseteq \{1,\dots, N\}$. Upon normalization, this yields the identity 
\[
    \frac{1}{\mathbb{V}[\QoI]} \sum_{J \subseteq \{1,\dots, N\}} V_J = 1.
\]
The main Sobol index $S_n$ associated with the $n$-th parameter, $n=1,\dots, N$, measures the fraction of the variance generated by the singleton $J = \{n\}$, i.e.
\[
S_n := \frac{V_{\{n\}}}{\mathbb{V}[\QoI]}.
\]
The total Sobol index $S_n^T$ associated with the $n$-th parameter measures instead the fraction of the variance generated by the interaction of the $n$-parameter with any other subset of parameters
\[
S_{n}^T := \frac{1}{\mathbb{V}[\QoI]} \sum_{n \in J \subseteq \{1,\dots, N\}} V_J \geq S_n.
\]
%\lorenzo{Note therefore that $\sum_{n=1}^NS_n \leq 1$, since such sum neglects all interaction terms, whereas  $\sum_{n=1}^NS_n^T \geq 1$ since interaction terms are counted in multiple $S_n^T$; for instance, $V_{j,k,l}$ (with $1 \leq j,k,l \leq N$ and $j \neq k \neq l$) contributes to $S_j^T,S_k^T,S_l^T$.}
The Sparse Grids Matlab Kit provides a dedicated routine for computing both main and total Sobol indices by postprocessing the sparse-grids surrogates \cite[Section 4.5]{Piazzola.Tamellini:24}.

\paragraph{Morris indices.} This technique roughly speaking measures the sensitivity of QoI with respect to a parameter $y_n$ by computing
an average over the parameter space of the partial derivatives $\partial_{y_n}\QoI$ (approximated by finite differences),
and can thus be seen as a global-local sensitivity measure.
More precisely, let $\{\yy_p\}_{p=1}^P \subseteq \Gamma$ be $P \geq 1$ random samples from the set of all admissible parameter combinations.
For $h_1, h_2, \ldots, h_N > 0$ and for all $p = 1,\dots, P$, consider the finite sequence
\[
\yy_{p,0} := \yy_{p}  \; \mapsto \; \yy_{p,1} := \yy_{p,0} + h_1 \ee_1 \; \mapsto \; \dots \; \mapsto \; \yy_{p,N} := \yy_{p, N-1} + h_N \ee_N
\]
where $\ee_1, \dots, \ee_N$ are the canonical unit basis vectors of $\mathbb{R}^N$. Assuming that each point in the sequence is still an admissible parameter combination, i.e. it is still in $\Gamma$, the elementary factor is  
\begin{equation}\label{eq:morris-factor}
    EE_{n,p} := \frac{\QoI(\yy_{p,n}) - \QoI(\yy_{p,n-1})}{h_n}, \qquad n=1,\dots, N.
\end{equation}
Then, we defined the Morris index associated with the $n$-th parameter by averaging the corresponding elementary factors 
\begin{equation}\label{eq:morris}
    m_n := \frac{1}{P} \sum_{p=1}^{P} \big| EE_{n,p} \big|. 
\end{equation}
Note that the absolute value prevents from cancellation effects. In analogy with Sobol indices, we compute Morris indices by postprocessing the sparse-grids surrogates.

\paragraph{Comparison.} Upon having computed Sobol and Morris indices, we check their agreement, to robustify the findings of the sensitivity analysis.
To this end, it is worth noticing that Sobol indices are dimensionless, normalized quantities, whereas Morris indices are dimensional numbers. For a fair comparison, we consider the following normalized Morris indices 
\begin{equation*}
\label{eq:morris-normalized}
\overline{m}_n := \frac{m_n}{\sum_{i=1}^N m_i}.
\end{equation*}
%\sidenotelorenzo{qui sarebbe buona cosa leggere le referenze canoniche \cite{Saltelli:08, Saltelli.Tarantola.Campolongo.Ratto:04, Iooss.Lemaitre:15}, che hanno sicuramente un po' di insight sul confronto, e forse anche queste che aveva trovato Vittoria \cite{Weber.Theers.Surmann.Ligges.Weihs:18,Feng.Lu.Yang:19}.}

% ----------------------------------------------------- %
%                      INVERSION
% ----------------------------------------------------- %

\subsection{Bayesian inversion}
\label{sec:bayes}
We resort to Bayesian techniques %\removecite{BuiThanh.Ghattas.Martin.Stadler:13,Petra.Martin.Stadler.Ghattas:14,Stuart:10}
\cite{BuiThanh.Ghattas.Martin.Stadler:13,Petra.Martin.Stadler.Ghattas:14} to derive an expression of the posterior pdf and to sample it.
According to Bayes theorem for conditional probabilities
\begin{equation}
  \label{eq:bayes-post}
  \rho_{\text{post}}(\yy) = \frac{1}{C} \mathcal{L}(M_1^*,\ldots,M_K^*; \yy) \rho_{\text{prior}}(\yy), \qquad \text{for $\yy \in \Gamma$}
\end{equation}
where $C$ is a normalization constant and $\mathcal{L}$ denotes the likelihood function. The latter is defined as
\begin{equation}
  \label{eq:likelihood}
  \mathcal{L}(M_1^*,\ldots,M_K^*; \yy) := \prod_{k=1}^K \rho_{\varepsilon_k}(M^*_k - M(\yy, t_k))
\end{equation}
where $\rho_{\varepsilon_k}$ is the pdf of the noise $\varepsilon_k$ from \eqref{eq:data_model} -- in our case, is a Gaussian distribution for $k=1,\ldots, K$.
Roughly speaking, evaluating the likelihood in $\yy$ measures how much plausible it is that the data $(M^*_1,\dots, M^*_K)$
were generated by $\yy_{\text{true}} = \yy$. Crucially, we remark that:
\begin{itemize}
\item evaluating the likelihood at $\yy$ requires running the HDG solver
  (or, more conveniently, evaluating the sparse-grid surrogate) to obtain the values of $M(\yy, t_k)$;
\item being able to evaluate the product $\mathcal{L} (M_1^*,\ldots,M_K^*; \yy) \rho_{\text{prior}}(\yy)$,
  i.e., being able to evaluate the posterior distribution up to the multiplicative constant $C$,
  is enough for move forward with our discussion.
\end{itemize}
% it is enough to be able to evaluate the posterior distribution up to the multiplicative constant $C$,
% Owing to \eqref{eq:data_model}, the pdf $\rho_{\varepsilon_k}$ is a Gaussian distribution for $k=1,\ldots, K$.
%Thus, the posterior distribution can be approximately evaluated, up to the multiplicative constant $C$, via \eqref{eq:bayes-post}-\eqref{eq:likelihood}, upon approximating the QoI $M$. To this end, using surrogates rather than the HDG solver makes the overall evaluation of $\rho_{\text{post}}$ much less computationally demanding. 

In particular, evaluating the posterior distribution up to $C$ is sufficient to generate
a sample distributed according to $\rho_{\text{post}}$, which will be needed in the subsequent forward UQ analysis (third step of the UQ workflow).
We consider in this work both a more simplistic Gaussian approximation and a more advanced Markov-Chain Monte-Carlo algorithm. For both techniques, it is useful introducing the Maximum-A-Posteriori (MAP) estimate $\yy_{\text{MAP}}\in \Gamma$. The latter is defined as the parameter combination that most likely generates the data $(M_1^*, \dots, M^*_K)$:
\begin{equation*}
\label{eq:MAP-definition}
\yy_{\text{MAP}} := \argmax_{\yy \in \Gamma} \rho_{\text{post}}(\yy) = \argmin_{\yy \in \Gamma} \left( - \log \rho_{\text{post}}(\yy) \right).
\end{equation*}
The second equivalent definition follows from the monotonicity of the logarithm and it is classically adopted in literature for the sake of stability of its numerical approximation.
We combine this formula with \eqref{eq:prior-pdf}, \eqref{eq:bayes-post} and \eqref{eq:likelihood}. Then, simple manipulations reveal
\begin{equation}
\label{eq:MAP}
\yy_{\text{MAP}} = \argmin_{\yy \in \Gamma} \sum_{k=1}^K \left( M_k^* - M(\yy, t_k) \right)^2,   
\end{equation}
which reveals the connection between Bayesian inversion and classical least squares calibration of parameters.%
\footnote{Tikhonov regularization terms for calibration can also be recovered in a Bayesian inversion framework by assuming a Gaussian prior for the uncertain parameters.}

Once $\yy_{\text{MAP}}$ is available, an estimate of the standard deviation $\sigma_\varepsilon$ of the measurement noise can be derived by its sample estimate:
\begin{equation}\label{eq:sigma-noise-sample-est}
\widetilde \sigma^2_{\varepsilon} := \frac{1}{K} \sum_{k=1}^K \left( M_k^* - M(\yy_{\text{MAP}},t_k) \right)^2.  
\end{equation}
Note that this estimate automatically incorporates also model errors
(in case the data are actually coming from real experiments) as well as the error in the approximation of $M$.
So, if sufficiently many data are available and a clear minumum has been identified,
it is advisable using $\widetilde \sigma_\varepsilon$ for the evaluation of $\rho_{\text{post}}$ and in \eqref{eq:GA-covariance} below
even when $\sigma^2_{\varepsilon}$ is explicitly known, as in our case.
After this preparation, we are in position to discuss the two above-mentioned techniques to sample the posterior pdf.

\paragraph{Gaussian approximation (GA).} This technique builds upon the assumption that $\rho_{\text{post}}$ is closed to a Gaussian distribution. We can expect that this property holds true if, e.g., we have 
\begin{equation*}
\label{eq:GA-assumption}
K \gg 1 \qquad \text{and} \qquad \sigma_\varepsilon \ll 1
\end{equation*}
in \eqref{eq:data_model}, meaning that several data are available with small noise, cf. \cite{BuiThanh.Ghattas.Martin.Stadler:13}.
%\removecite{BuiThanh.Ghattas.Martin.Stadler:13,Piazzola.Tamellini.Tempone:21} 
We approximate the posterior pdf by the multivariate Gaussian distribution with mean $\yy_\text{MAP}$ and covariance matrix
\begin{equation}
\label{eq:GA-covariance}
\Sigma_{\text{post}} := \widetilde \sigma^2_{\varepsilon} \left( J_M^T J_M \right)^{-1}
\end{equation}
with $\widetilde \sigma^2_{\varepsilon}$ from \eqref{eq:sigma-noise-sample-est}
and $J_M$ denoting the $K \times N$ Jacobian matrix
\begin{equation}
\label{eq:GA-Jacobian}
\left[J_M\right]_{k, n} := \partial_n M(\yy_\text{MAP}, t_k).
\end{equation}
Here $\partial_n$ indicates the derivatives of $M$ with respect to the $n-th$ parameter, that we approximate by simple finite differences applied to the sparse-grid surrogate model.
The expression for $\Sigma_{\text{post}}$ in \eqref{eq:GA-covariance} can be derived as an approximation of the inverse of the Hessian
of the least squares functional under minimization in (\ref{eq:MAP}), and thus represents a way of measuring the ``confidence'' in the MAP estimate
(if the minimum is a ``narrow valley'' of the functional under minimization, i.e., we are ``very confident'' about the MAP,
the Hessian matrix will have very large eigenvalues and correspondingly the posterior computed with said formula will have very small variances).

Once the covariance matrix is computed, the sample distributed according to $\rho_{\text{post}}$, for the subsequent forward UQ analysis
can be generated in a straightforward way, by employing standard sample generation for multivariate Gaussian distributions.

\paragraph{Markov Chain Monte Carlo (MCMC).} This class of algorithms does not assume a specific distribution for $\rho_{\text{post}}$. MCMC algorithms are iterative algorithms generating a \emph{chain} of $B$ samples $\mathcal{C}_B = \{\yy_{i}\}_{i=1}^B$, asymptotically distributed as $\rho_{\text{post}}$ and hence directly usable for the the subsequent forward UQ analysis. We use $\yy_{\text{MAP}}$ as the starting point of the chain; the next samples are generated depending on the specific MCMC algorithm in use, based on the evaluation of $\log(\rho_\text{post})$ via \eqref{eq:bayes-post} up to the multiplicative constant $C$. 

Among the different MCMC algorithms available in the literature, we use the slice sampling MCMC algorithm \cite{Neal:03},
which roughly speaking iteratively computes approximate counter-images of upper level sets of $\log(\rho_\text{post})$ and generates samples from such
counter-images. Despite the complexity of computing approximate counter-images, we prefer this algorithm 
over the more popular Metropolis--Hastings algorithm \cite{Petra.Martin.Stadler.Ghattas:14} %\removecite{Stuart:10,Petra.Martin.Stadler.Ghattas:14} 
because it does not require coming up with a so-called proposal distribution, which can be seen as a critical ``hyperparameter''
on whose choice relies the computational efficiency of the method.
% \sidenotelorenzo{Pietro chiedeva più dettagli, così forse va bene. Probabilmente vale la pena aggiungere anche i valori degli iperparametri di Slice Sampleing}

Regardless of the specific MCMC algorithm employed, we remark that it can happen that the autocorrelation of the samples of the chain $\mathcal{C}_B$ decays slowly in terms of the lag, which lowers the quality of the chain (i.e., whether the distribution of the samples in $\mathcal{C}_B$ is a good approximation of $\rho_{\text{post}}$). This issue can be mitigated by discarding $t-1$ samples after each selected one (\emph{thinning}). Moreover, one usually discards the first $b$ samples of the chain (\emph{burn-in}) prior to the \emph{thinnig} phase to mitigate the above-mentioned fact that the chain is only \emph{asymptotically} distributed as $\rho_{\text{post}}$, such that the final chain consists of only $(B-b)/t$ samples.

% We remark that the generation of each sample requires (at least) one evaluation of $M$. Thus, the generation of large samples by MCMC algorithms is much more computationally demanding than by GA, as the latter technique involves the evaluation of $M$ only via the definition of the covariance matrix. 

\paragraph{Concentration measures.} The marginal posterior pdf of each parameter is expected to be more concentrated than the prior one. The concentration of a pdf with mean $\mu$ and standard deviation $\sigma$ can be quantified via the Coefficient of Variation (CV), defined as
\begin{equation}
\label{eq:CV}
\mathrm{CV} := \frac{\sigma}{|\mu|}.
\end{equation}
We expect the coefficient of variation of the posterior pdf to be smaller than the one of the prior pdf. Moreover, when the prior and the posterior means are close, the difference in the coefficients of variation mostly hinges on the ratio of the respective standard deviations $\sigma_\text{prior}$ and $\sigma_\text{post}$. Therefore, the uncertainty reduction can be measured via the Concentration Factor (CF), defined as
\begin{equation}
\label{eq:CF}
\mathrm{CF} := \frac{\sigma_\text{post}}{\sigma_\text{prior}}.
\end{equation}

% ----------------------------------------------------- %
%                      FORWARD UQ
% ----------------------------------------------------- %

\subsection{Forward UQ}
\label{sec:forward-UQ}

Having a sample of $\rho_{\text{post}}$ at hand, we can finally perform the forward uncertainty quantification analysis of $I$, i.e., we propagate the residual uncertainty encoded in $\rho_{\text{post}}$ from the parameters to the QoI. To this end, we first evaluate $I(\yy, t)$ for $\yy$ in the sample of $\rho_{\text{post}}$ and $t \in [0,T]$. Once more, it is convenient approximating $I$ via the surrogates instead of the HDG solver. Then, we compute statistics if $I$, like mean and standard deviation, and we approximate the pdf of $I$ by a classical kernel density estimator. % \removecite{Parzen:62}.
We quantify the uncertainty reduction in comparison with the results obtained by propagating the prior uncertainty on the parameters \eqref{eq:prior-pdf} to $I$. To this end, we compute the concentration measures \eqref{eq:CV} and $\eqref{eq:CF}$ for $t \in [0, T]$.

%------------------------------------------------------------------------------------------------------
\section{Numerical results}\label{sec:numerical-results}
%------------------------------------------------------------------------------------------------------

In this section we first validate both the HDG solver introduced in Section~\ref{sec:HDG}
and the construction of suitable sparse-grids surrogates for $M$ and $I$. This preliminary step is instrumental to balance accuracy and cost in the successive computations. Then, we perform the three steps of the UQ workflow \eqref{eq:UQ-workflow}, according to the guidelines in Section~\ref{sec:UQ}, and we present our results.

All tests were implemented in Matlab and run on a standard laptop
powered by a Intel Core i7-1355U processor
and equipped with 16 GB of RAM. The HDG solver is in-house, based on \cite{Bertoluzza.Bretti.Pennacchio.Prudhomme:25+}. For all experiments, we use $N_\mathtt{t}$ steps of equal size in the time discretization and $N_\mathtt{s}$ elements of equal size in the space discretization. The sparse-grids surrogates are constructed and evaluated via the Sparse Grids Matlab Kit release 23-5  \cite{Piazzola.Tamellini:24}. %\removecite{Piazzola.Tamellini:24,Piazzola.Tamellini:23} %\sidenotelorenzo{add github sgmk?}
% All source codes are publicly available at the public repository \sidenotelorenzo{togliere}
% \begin{center}
% \url{https://github.com/lorenzo-tamellini/UQ4KS.git/}    
% \end{center}

\subsection{HDG solver validation}
\begin{figure}[tp]
  \centering
  \includegraphics[width=0.4\linewidth]{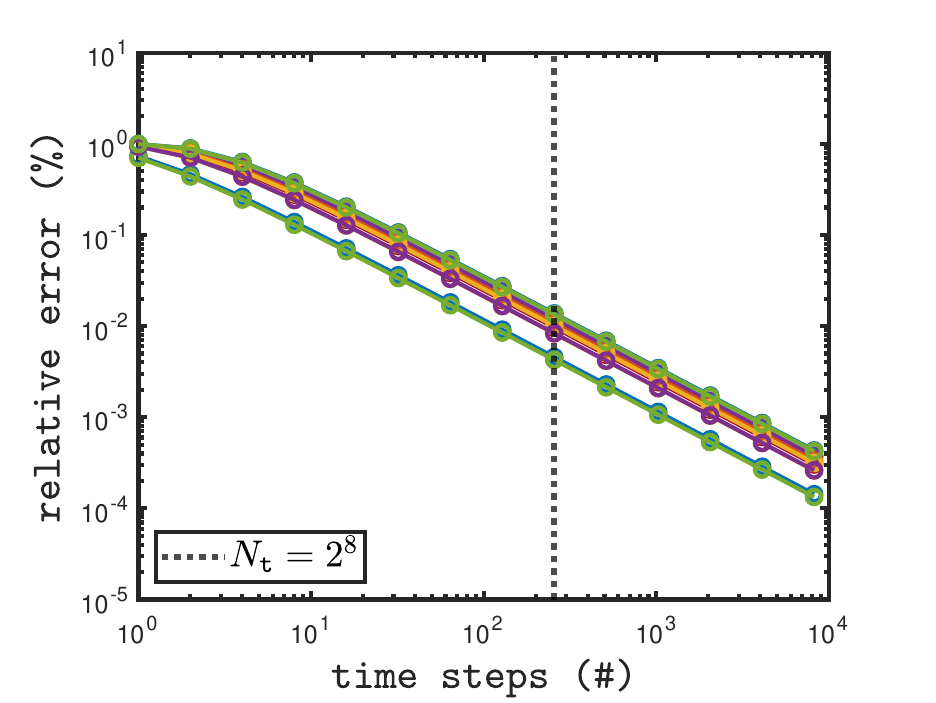}\qquad
  \includegraphics[width=0.4\linewidth]{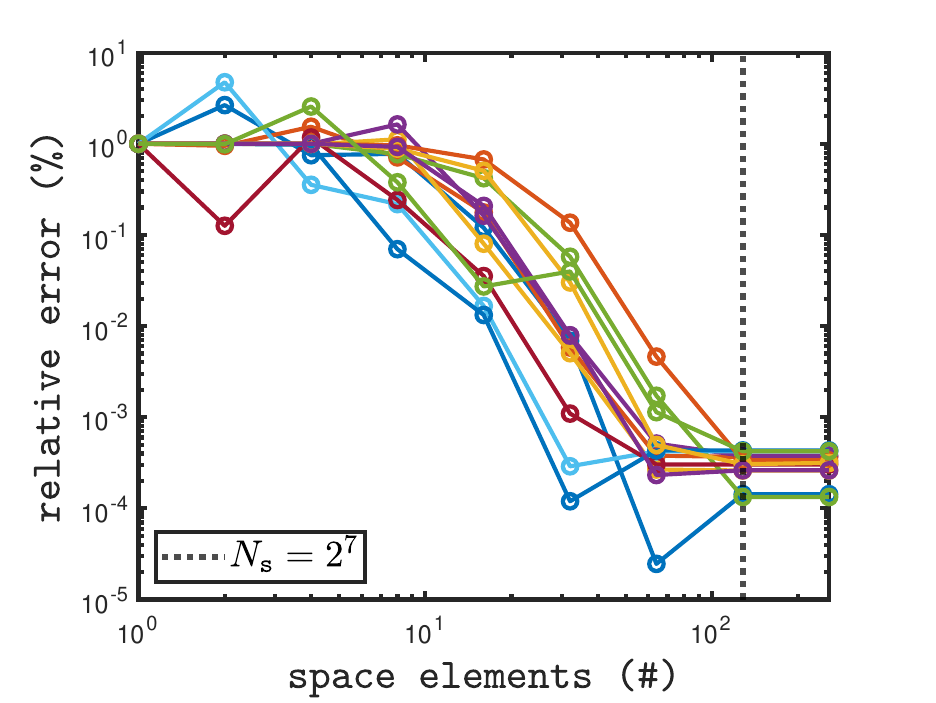}
  \caption{Plot of $\mathtt{err}(\yy, N_\mathtt{t}, N_\mathtt{s})$ for $12$ random samples $\yy \in \Gamma$ (one line per sample). Left: error versus $N_\mathtt{t}$ with fixed $N_\mathtt{s} = 2^8$. Right: error versus $N_\texttt{s}$ with fixed $N_\mathtt{t} = 2^{13}$. The vertical dotted lines indicate the values of $N_\mathtt{t}$ and $N_\mathtt{s}$ fixed in \eqref{eq:HDG-solver-hyperparameters} for the subsequent analysis.}
  \label{fig:test_solver}
\end{figure}

%
% \sidenotepietro{Riportare i valori utilizzati per i parametri di stabilizzazione}
To assess the accuracy of the HDG solver, we observe that the closed formula 
\begin{equation}
\label{eq:I-formula}
I(\yy, t) = \frac{ \kappa_\varphi(e^{-a t} - e^{-\rho t})}{\rho - a}\left( \int_0^L \exp\left(-\frac{(x - \ c_{\varphi})^2}{2\sigma_{\varphi}^2}\right) dx \right)^{-1}
\end{equation}
can be inferred by integrating the second equation of \eqref{eq:modello2eq} in space, exploiting the boundary conditions for $\varphi$ and solving the resulting ordinary differential equation. We set the penalty parameters in \eqref{eq:HDG-numerical-fluxes} as $\tau_{u} = \tau_{\varphi}=1$. Then, we consider the relative error
\begin{equation*}
\label{eq:error-I}
\mathtt{err}(\yy, N_\mathtt{t}, N_\mathtt{s}) := \max_{n=1,\dots N_{\mathtt{t}}}\frac{|I(\yy, t_n) - I_\mathrm{HDG}(\yy, t_n)|}{I(\yy, t_n)}
\end{equation*}
where $I_\mathrm{HDG}$ is the approximation of $I$ returned by the solver and $(t_n)_{n=0}^{N_\mathtt{t}}$ are the points in the time discretization.
%\sidenotelorenzo{la validazione vs il centro di massa si riesce a fare a questo punto? eventualmente con ref numerica. PZ: in teoria si può fare, ma ci vogliono molti più intervalli in spazio e tempo ed io ho già spinto il mio PC piuttosto in là. L'idea comunque è che se approssimi una QoI con le velocità di convergenza attese in spazio e tempo, allora tendenzialmente stai approssimando tutta la soluzione con quelle velocità (oppure sei molto, molto fortunato)} 
We account for the variability of the error with respect to $\yy$ by generating $12$ random samples in the set $\Gamma$. For each sample, we investigate the error decay with respect to both the time and the space discretization. In the first case, we fix $N_\texttt{s} = 2^8$ and solve for $N_\texttt{t} = 2^0, 2^1,\dots, 2^{13}$. We observe first order convergence for all samples, as expected for the backward Euler scheme, cf. Figure~\ref{fig:test_solver} (left panel). In the other case, we fix $N_\mathtt{t} = 2^{13}$ and solve for $N_\mathtt{s} = 2^0, 2^1, \dots 2^8$. We initially observe stagnation, due to insufficient spatial resolution, compared to the data variability. Then superlinear convergence takes place (with more variability among samples than in the other experiment), followed again by stagnation, due to insufficient time resolution, cf. Figure~\ref{fig:test_solver} (right panel).

Owing to the results above, a convenient balance of accuracy and cost for the subsequent analysis is obtained by fixing 
\begin{equation}
\label{eq:HDG-solver-hyperparameters}
N_\mathtt{t} = 2^8 \qquad \text{and} \qquad N_\mathtt{s} = 2^7.
\end{equation}
We validate this choice by generating $60$ random samples in $\Gamma$. The maximum and the root mean square of \texttt{err} with respect to all samples are, respectively, 
\[
\texttt{max} = 1.40\% \qquad \text{and} \qquad \texttt{rms} = 0.98\%.
\]
Running the solver for all samples (without parallelization) takes approximately $130s$. These results and the subsequent ones are obtained with the penalization parameters $\tau_u = 1 = \tau_\varphi$ for the numerical fluxes \eqref{eq:HDG-numerical-fluxes}. In the Newton solver, we set the tolerance to $10^{-5}$ and perform a maximum of $10$ iterations. Such values appear to make the Newton solver error negligible in comparison to the above errors.

\subsection{Surrogate models validation}
\label{sec:surrogates-validation}
We exploit the HDG solver with $N_\mathtt{t}$ and $N_\mathtt{s}$ as in \eqref{eq:HDG-solver-hyperparameters} to construct sparse-grid surrogates for the QoIs $M$ and $I$. More precisely, we construct a surrogate for each QoI at each time step of HDG solver. For this purpose, we use the Sparse Grids Matlab Kit with the following standard setup: 
\begin{itemize}
    \item \emph{Clenshaw--Curtis collocation abscissae} for each uncertain parameters, see \cite[Section 3]{Piazzola.Tamellini:24};
    \item \emph{level-to-knots} function of type `doubling', see \cite[Eq. (5)]{Piazzola.Tamellini:24};
    \item  \emph{index set} of type `Total Degree', see \cite[Eq. (15), (17)]{Piazzola.Tamellini:24}.
\end{itemize}
This combination of hyperparameters results in a so-called Smolyak sparse-grid and it is tailored to uncertain parameters with uniform prior distributions, as prescribed by \eqref{eq:prior-pdf}.

\begin{figure}[tp]
  \centering
  \includegraphics[width=0.4\linewidth]{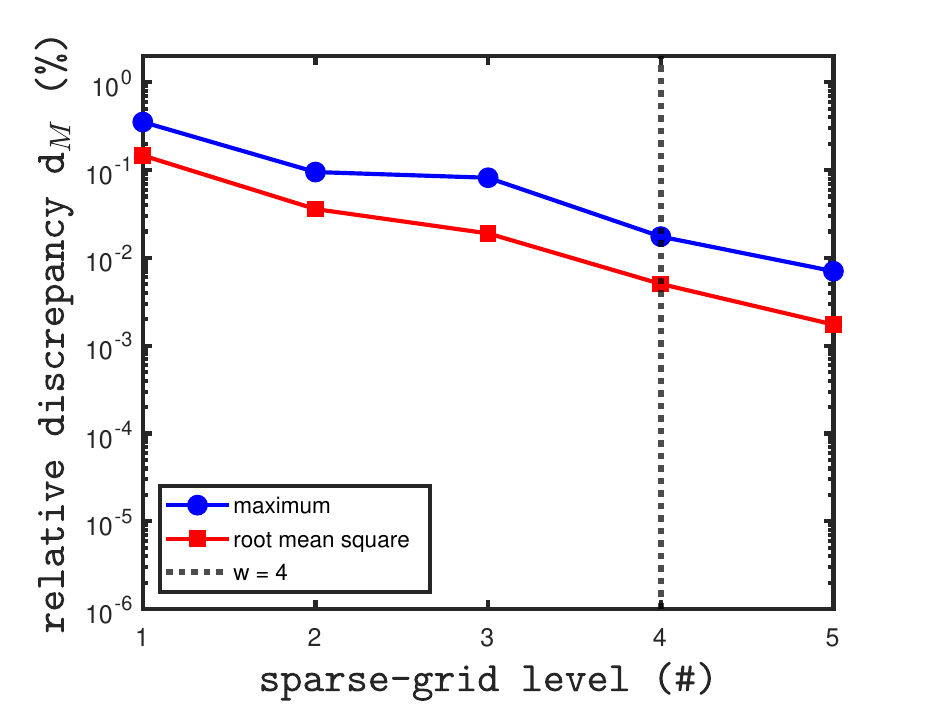}\qquad
  \includegraphics[width=0.4\linewidth]{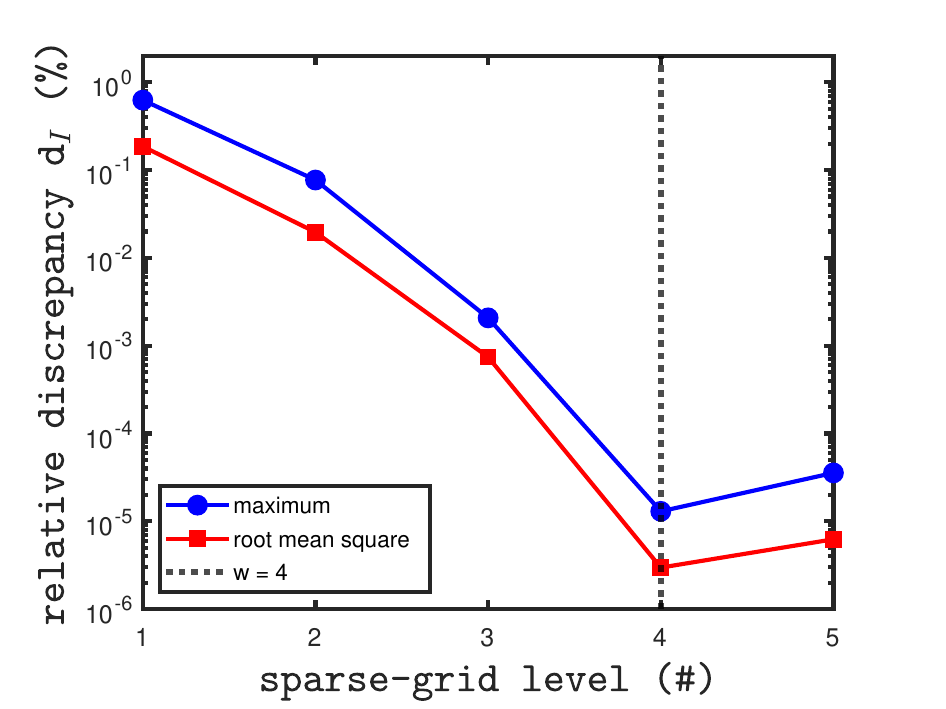}
  \caption{Maximum and root mean square of the relative discrepancy versus the sparse-grid level $w$ for the QoIs $M$ (left) and $I$ (right). The vertical dotted lines indicate the value of $w$ fixed in \eqref{eq:w-fixed} for the subsequent analysis.}
  \label{fig:test_surrogate}
\end{figure}

As mentioned in Section \ref{sec:surrogates}, the number of nodes in the sparse-grid, hence the accuracy of the surrogate models, is controlled by the sparse-grid level $w$, see Table~\ref{tab:w_sparsegrids}. For a proper selection of $w$, we generate $60$ random samples in $\Gamma$. For each sample, we compute the relative discrepancy between the quantities of interest $M_\mathrm{HDG}$ and $I_\mathrm{HDG}$ computed by the HDG solver and their approximations $M_{\mathrm{S}(w)}$ and $I_{\mathrm{S}(w)}$ obtained by the sparse-grid surrogate of level $w$
\[
\mathtt{d}_M(\yy, w) := \max_{k=1,\dots, N_\mathtt{t}} \frac{|M_{\mathrm{HDG}}(\yy, t_k) - M_{\mathrm{S}(w)}(\yy, t_k)|}{|M_{\mathrm{HDG}}(\yy, t_k)|}
\]
and 
\[
\mathtt{d}_I(\yy, w) := \max_{k=1,\dots, N_\mathtt{t}} \frac{|I_{\mathrm{HDG}}(\yy, t_k) - I_{\mathrm{S}(w)}(\yy, t_k)|}{|I_{\mathrm{HDG}}(\yy, t_k)|}.
\]
Both the maximum and the root mean square of $\mathtt{d}_I$, with respect to all samples, decay quickly to zero, see Figure~\ref{fig:test_surrogate} (right panel). For $\mathtt{d}_M$ (left panel) the convergence is less pronounced but still clear.

%\sidenotepietro{Mi lascia un po' perplesso la risalita della discrepanza in I. Ad esempio, l'analogo test con i 4 parametri selezionati per l'inversione non evidenzia la risalita. Magari rifare il test e approfondire. }

Owing to the results above, a convenient balance of accuracy and cost for the subsequent developments is obtained by fixing
\begin{equation}
\label{eq:w-fixed}
w = 4.
\end{equation}
In this case, the maximum and the root mean square with respect to the samples are, respectively
\[
\texttt{max} = 1.75\% \qquad \text{and} \qquad \texttt{rms} = 0.51\%
\]
for $\mathtt{d}_M$ and 
\[
\texttt{max} < 0,01\% \qquad \text{and} \qquad \texttt{rms} < 0.01\%
\]
for $\mathtt{d}_I$. Assessing the computational cost for the evaluation of the surrogates is not immediate,
because the dedicated routine is more efficient if it is called once for many samples rather than once per sample, due to the fact that a certain number of
operations must be performed regardless of the number of required evaluations.
For instance, considering the $60$ samples above, is takes about $0.15s$ to evaluate the
all the $60$ samples in one go vs about $0.60s$ to perform $60$ times the evaluation, one sample at a time (without parallelization).

%-----------------------------------------------
\subsection{UQ workflow step 1: sensitivity analysis}
\label{sec:sensitivity-numerics}
\begin{figure}[tp]
  \captionsetup[subfigure]{labelformat=empty}
  \centering
  \subfloat{\includegraphics[width=0.31\linewidth]{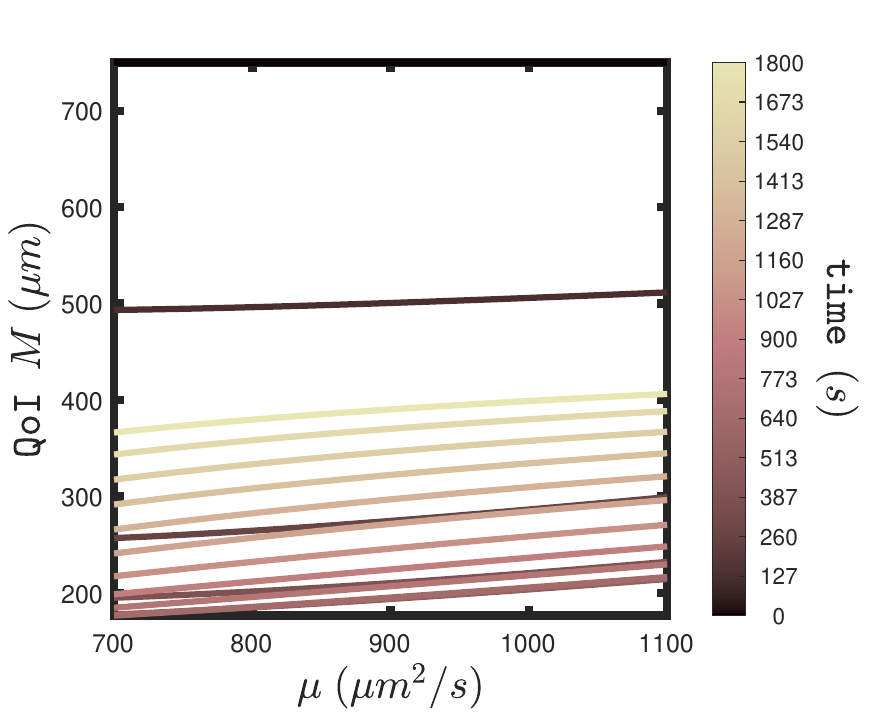}}  
  \quad
  \subfloat{\includegraphics[width=0.31\linewidth]{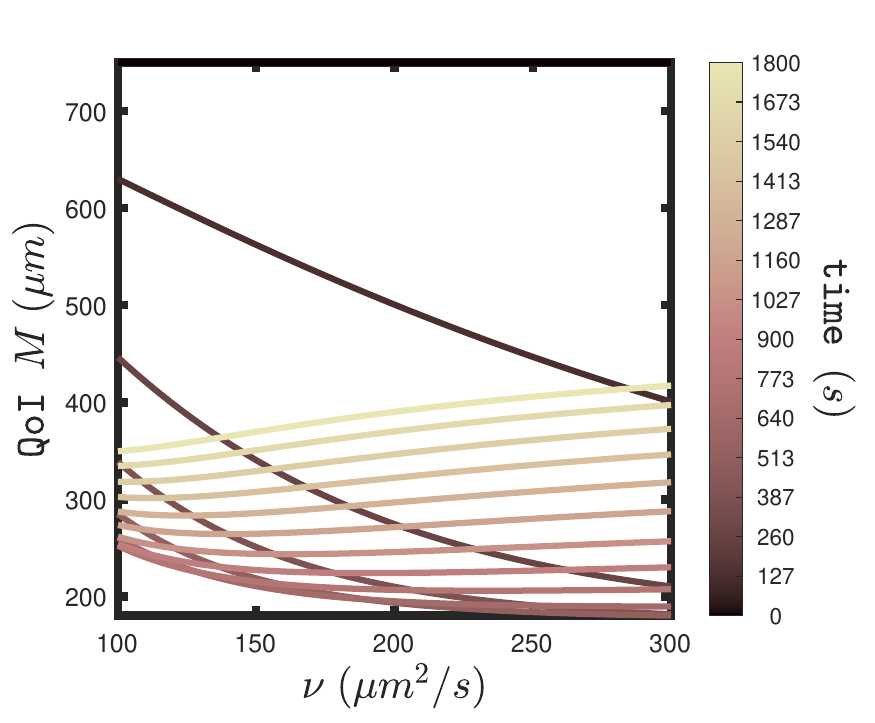}} 
  \quad
  \subfloat{\includegraphics[width=0.31\linewidth]{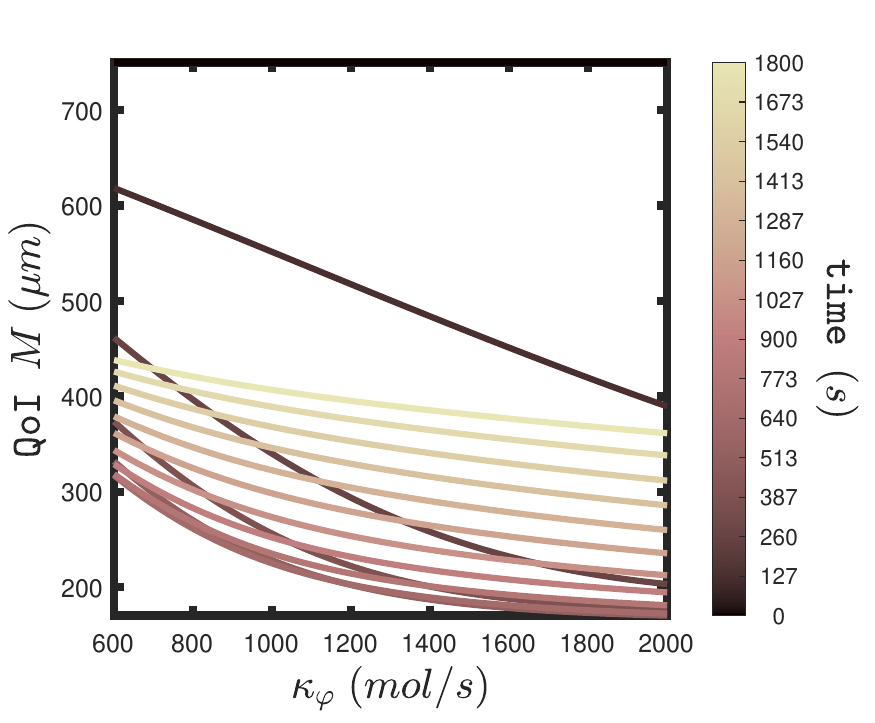}}\\
  \subfloat{\includegraphics[width=0.31\linewidth]{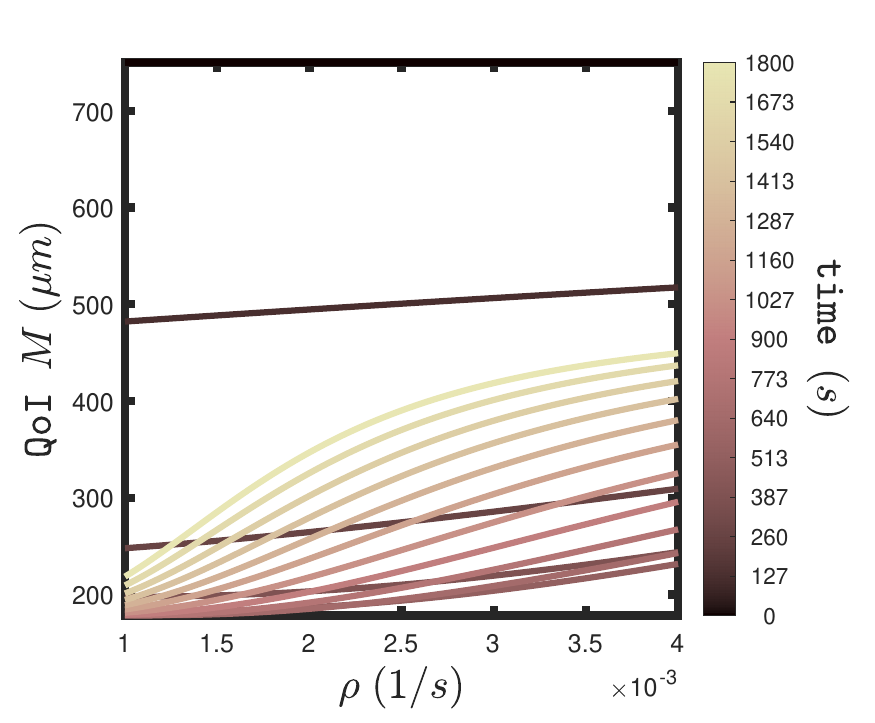}} 
  \quad
  \subfloat{\includegraphics[width=0.31\linewidth]{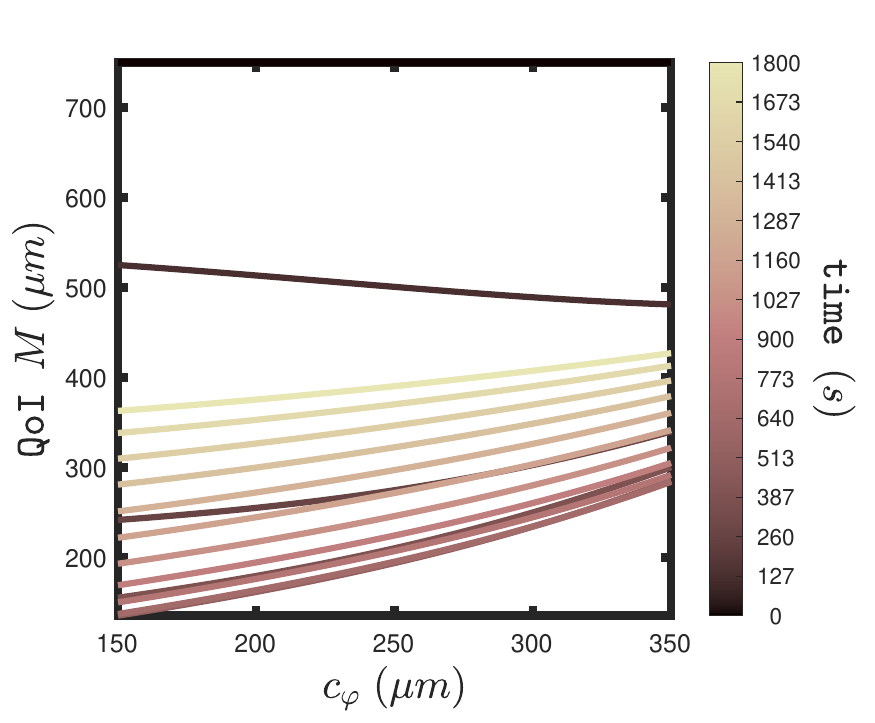}} 
  \quad
  \subfloat{\includegraphics[width=0.31\linewidth]{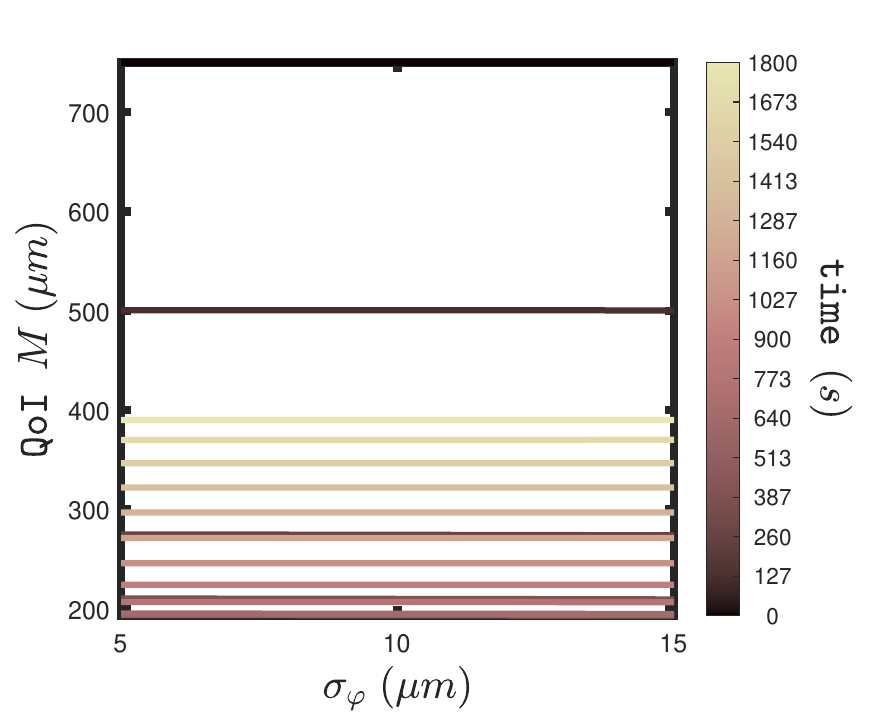}}
   \caption{Evolution in time of the response curves of $M$ with respect to the parameters.}
  \label{fig:response-median}
\end{figure}

\begin{figure}[tp]
  \captionsetup[subfigure]{labelformat=empty}
  \centering
  \subfloat{\includegraphics[width=0.31\linewidth]{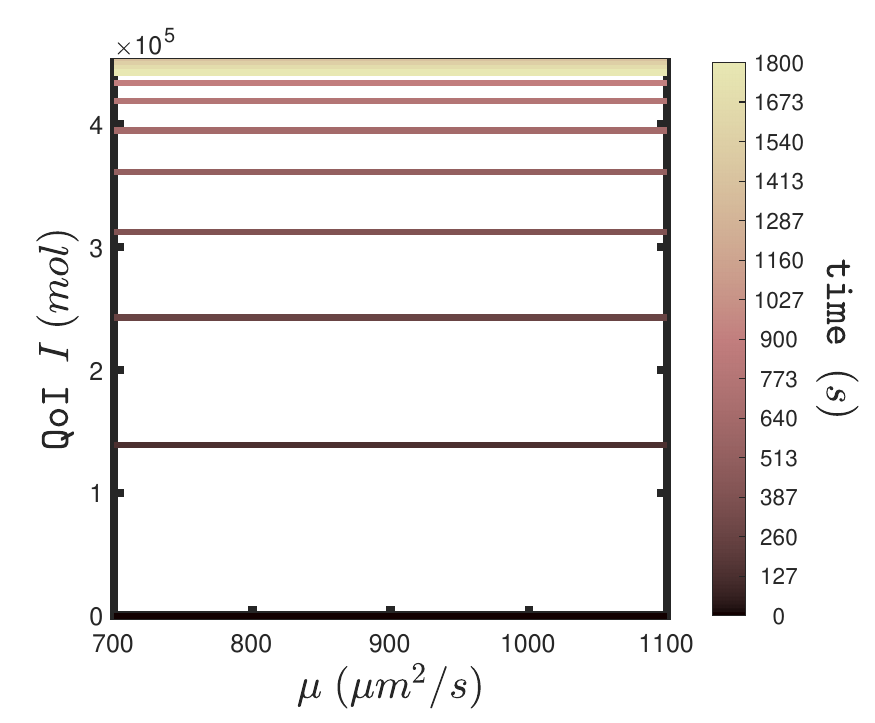}}  
  \quad
  \subfloat{\includegraphics[width=0.31\linewidth]{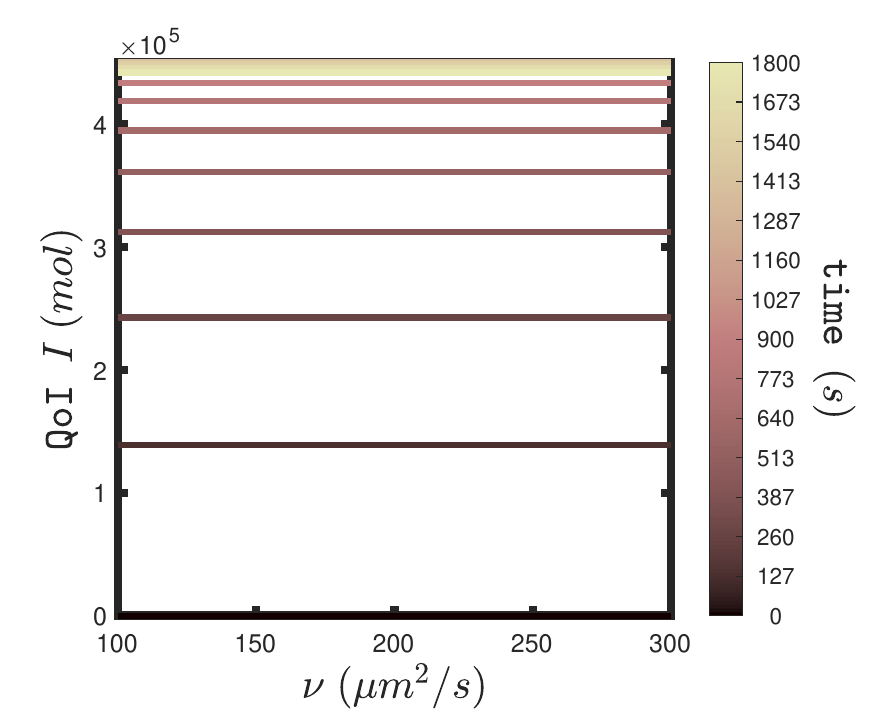}} 
  \quad
  \subfloat{\includegraphics[width=0.31\linewidth]{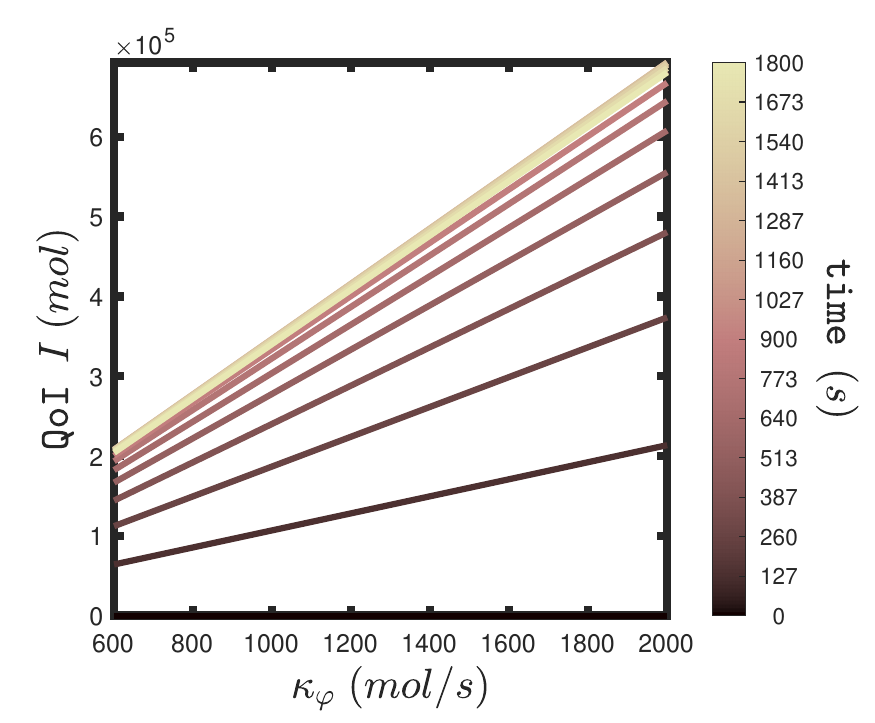}}\\
  \subfloat{\includegraphics[width=0.31\linewidth]{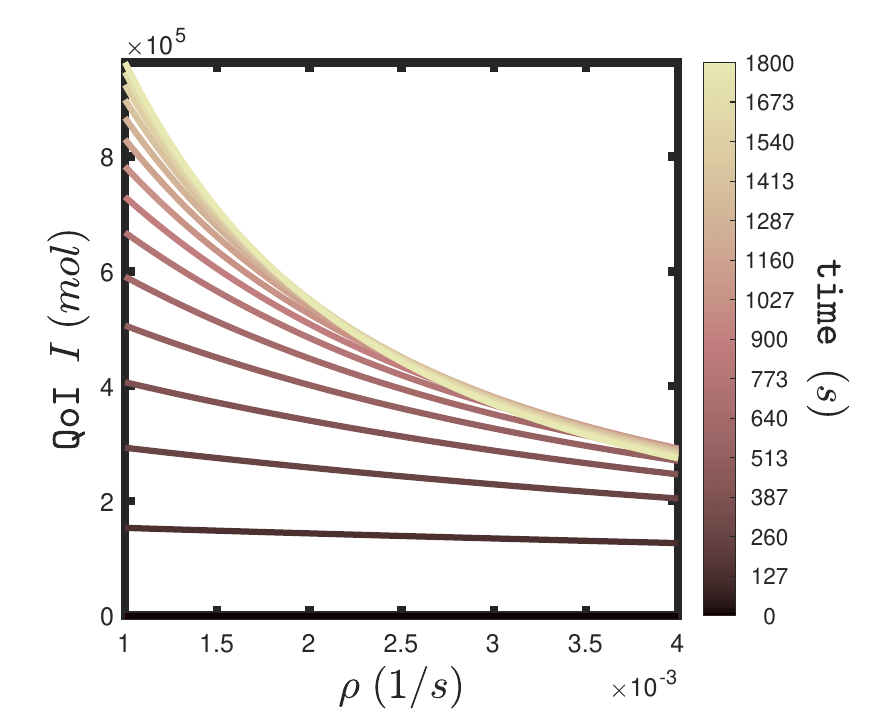}} 
  \quad
  \subfloat{\includegraphics[width=0.31\linewidth]{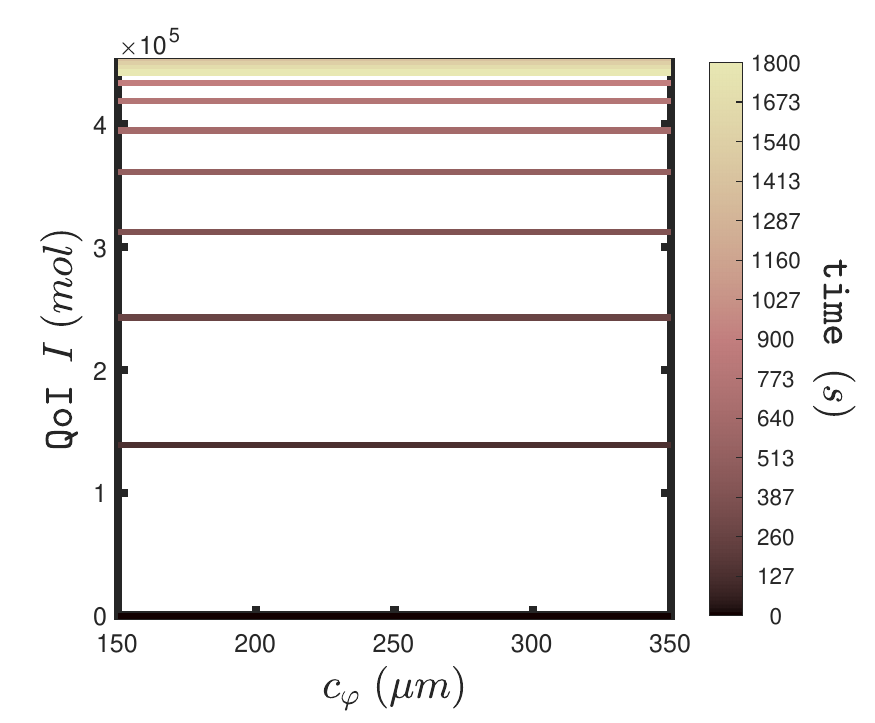}} 
  \quad
  \subfloat{\includegraphics[width=0.31\linewidth]{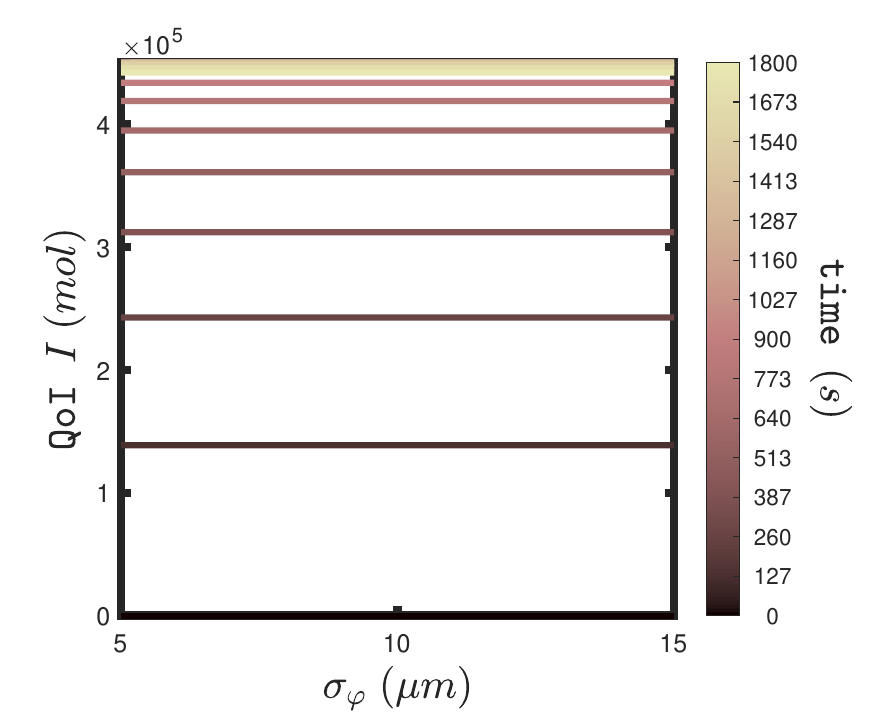}}
   \caption{Evolution in time of the response curves of $I$ with respect to the parameters.}
  \label{fig:response-integral}
\end{figure}

We use the surrogate models from Section~\ref{sec:surrogates-validation} for the sensitivity analysis of $M$ and $I$, as outlined in Section~\ref{sec:sensitivity}. We first plot some response curves of the QoIs at different time instants. As already hinted in Section \ref{sec:sensitivity}, the curves are obtained by letting one parameter vary in its range, whereas the other ones are fixed to the midpoint of the respective range, Table~\ref{tab:parametri} (one-at-a-time sensitivity analysis).
According to Figure~\ref{fig:response-median}, the parameters $\mu$ and $\sigma_\varphi$ play a negligible role for $M$, whereas the other ones appear to be substantially impactful. Figure~\ref{fig:response-integral} suggests that $I$ is insensitive to $\mu, \nu, c_\varphi, \sigma_\varphi$, whereas it is clearly responsive to the other parameters. 

\begin{figure}[tp]
  \captionsetup[subfigure]{labelformat=empty}
  \centering
  \subfloat[$(\mu)$]{\includegraphics[width=0.31\linewidth]{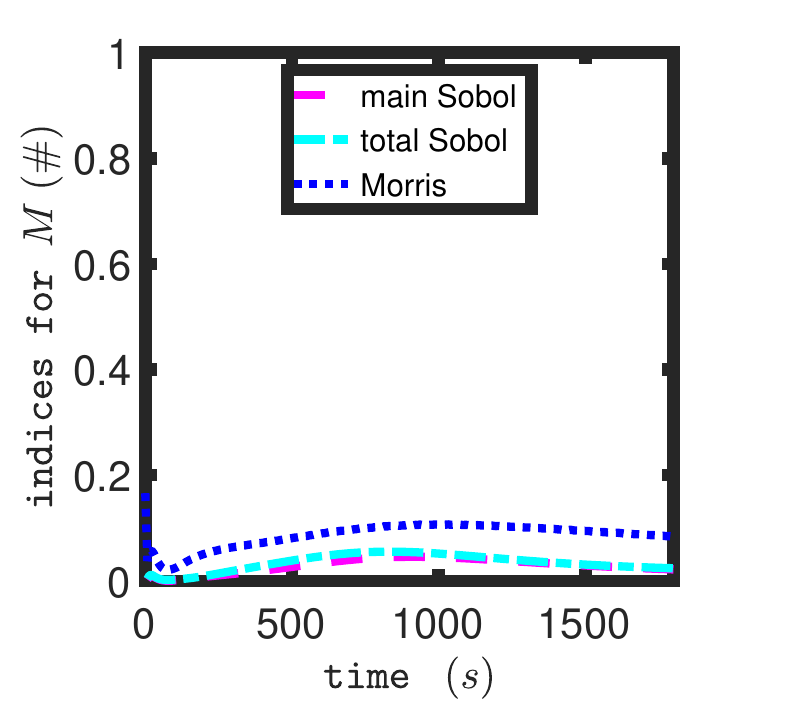}}  
  \quad
  \subfloat[$(\nu)$]{\includegraphics[width=0.31\linewidth]{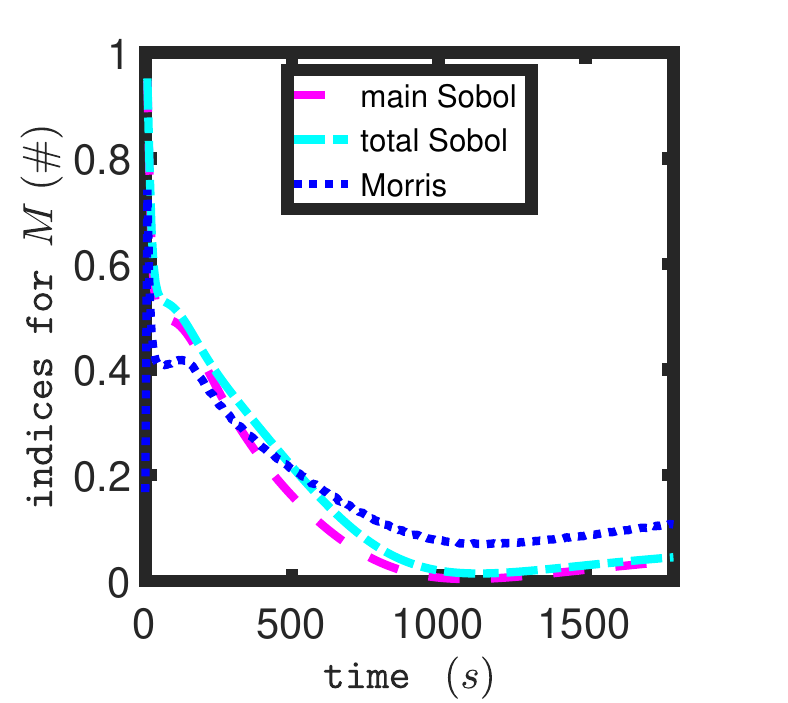}} 
  \quad
  \subfloat[$(\kappa_\varphi)$]{\includegraphics[width=0.31\linewidth]{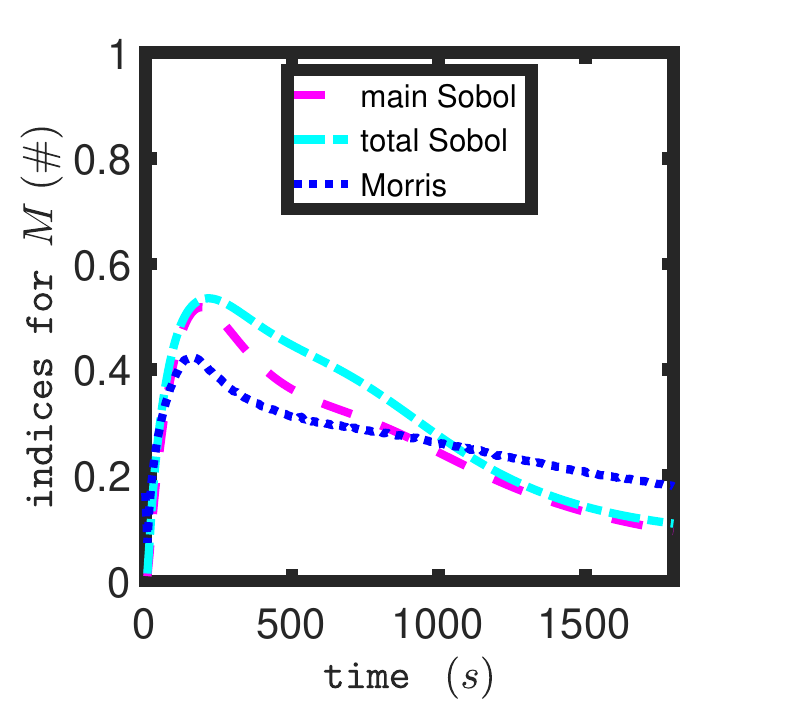}} \\
  \subfloat[$(\rho)$]{\includegraphics[width=0.31\linewidth]{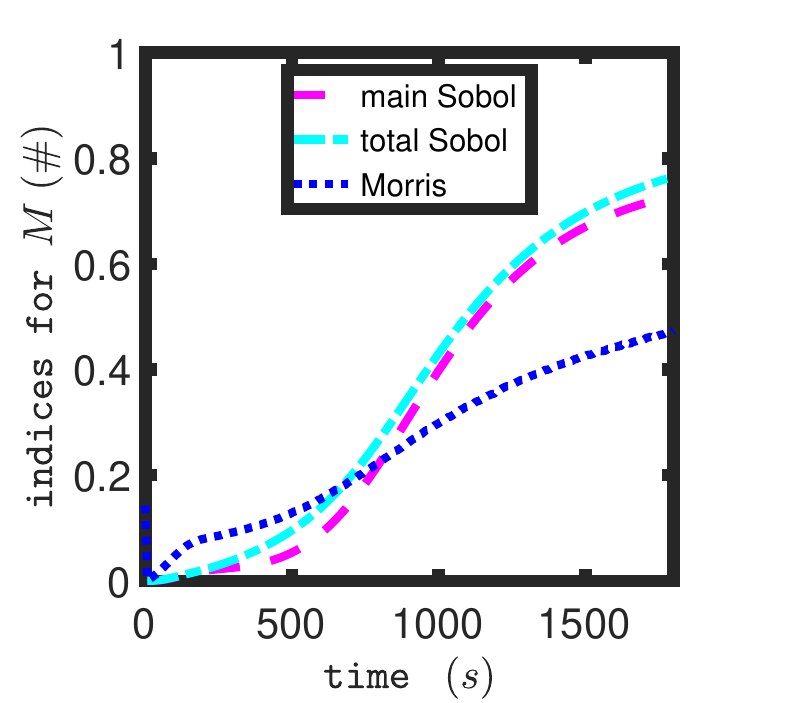}} 
  \quad
  \subfloat[$(c_\varphi)$]{\includegraphics[width=0.31\linewidth]{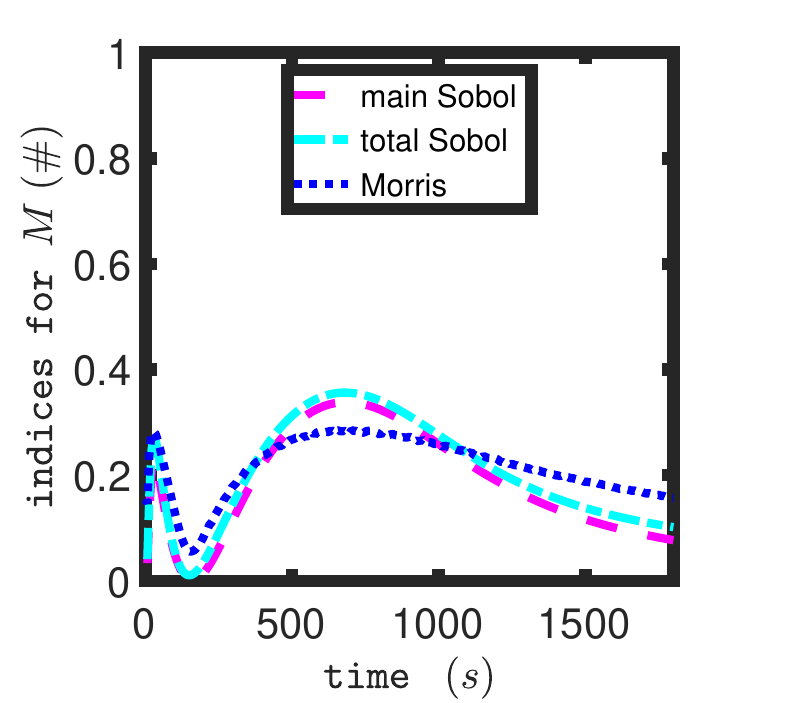}} 
  \quad
  \subfloat[$(\sigma_\varphi)$]{\includegraphics[width=0.31\linewidth]{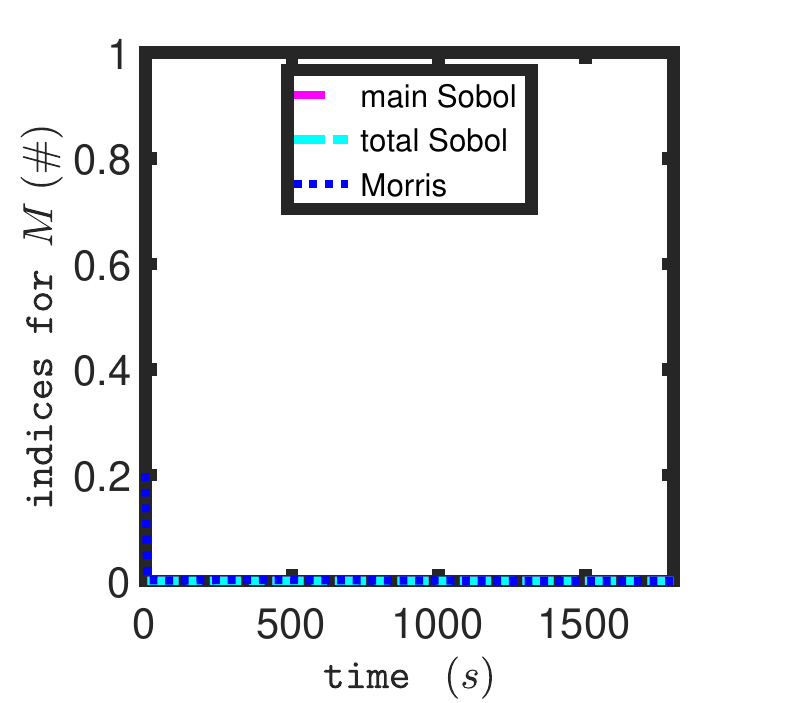}} 
   \caption{Evolution in time of the sensitivity indices of $M$.}
  \label{fig:indices-median}
\end{figure}

\begin{figure}[tp]
  \captionsetup[subfigure]{labelformat=empty}
  \centering
  \subfloat[$(\mu)$]{\includegraphics[width=0.31\linewidth]{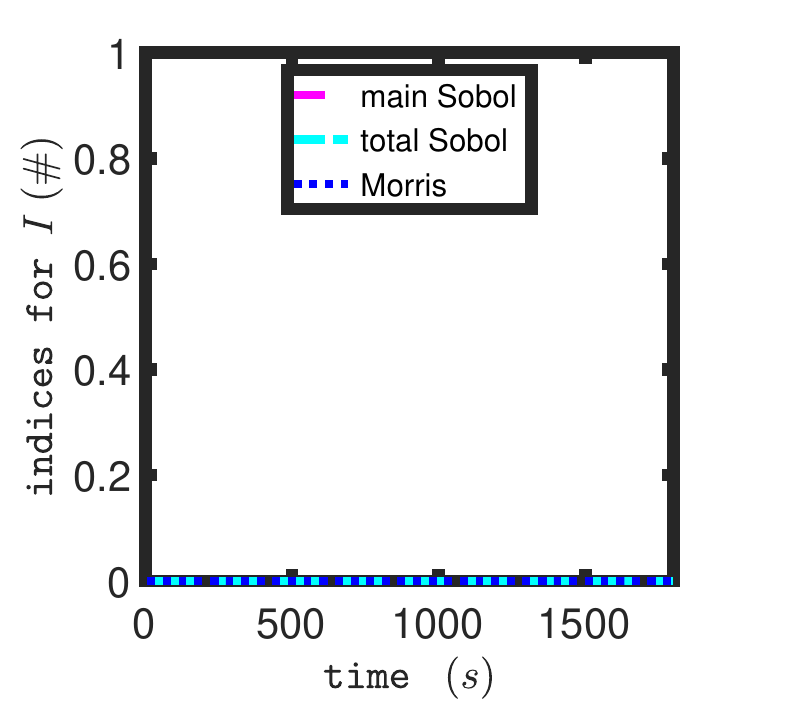}} 
  \quad
  \subfloat[$(\nu)$]{\includegraphics[width=0.31\linewidth]{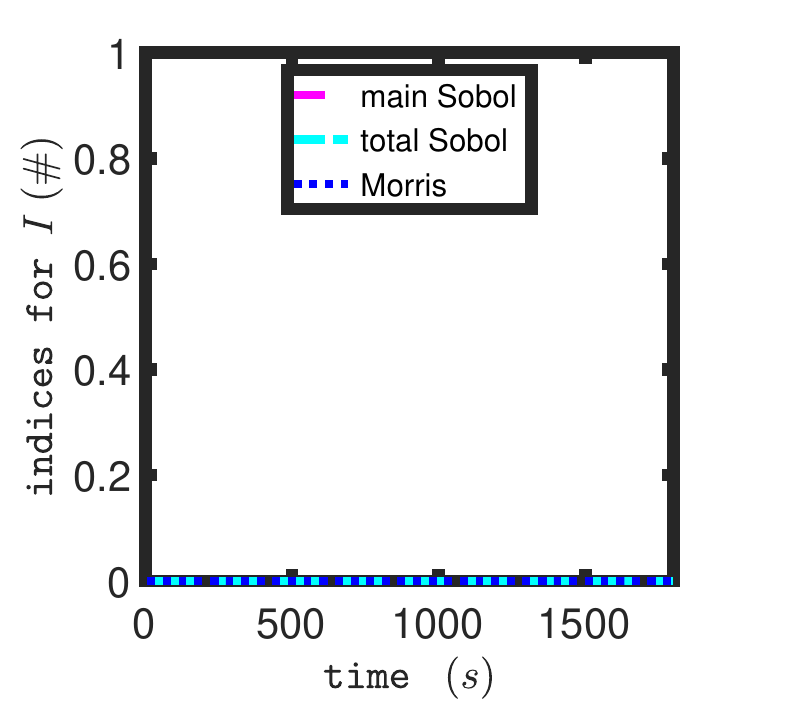}} 
  \quad
  \subfloat[$(\kappa_\varphi)$]{\includegraphics[width=0.31\linewidth]{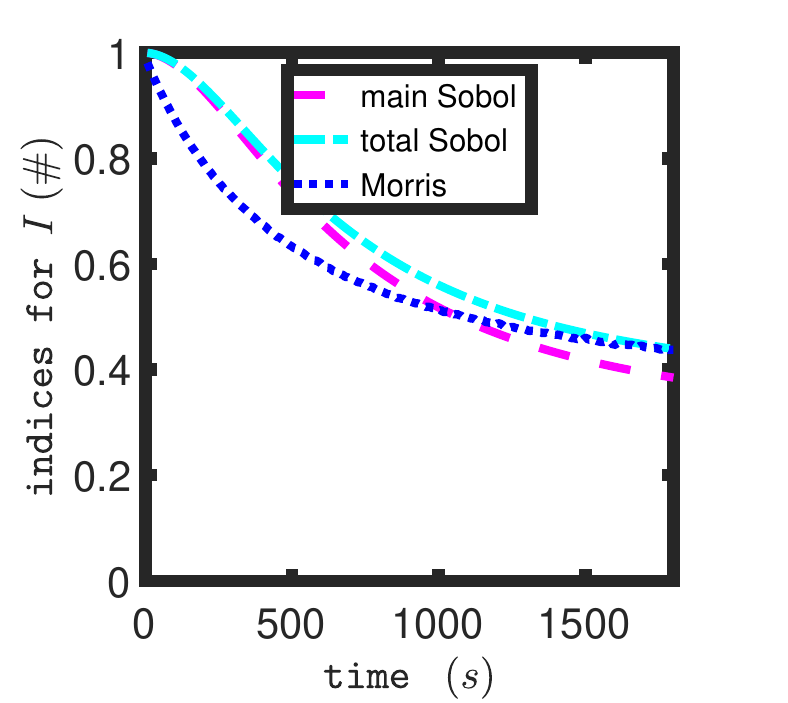}} \\
  \subfloat[$(\rho)$]{\includegraphics[width=0.31\linewidth]{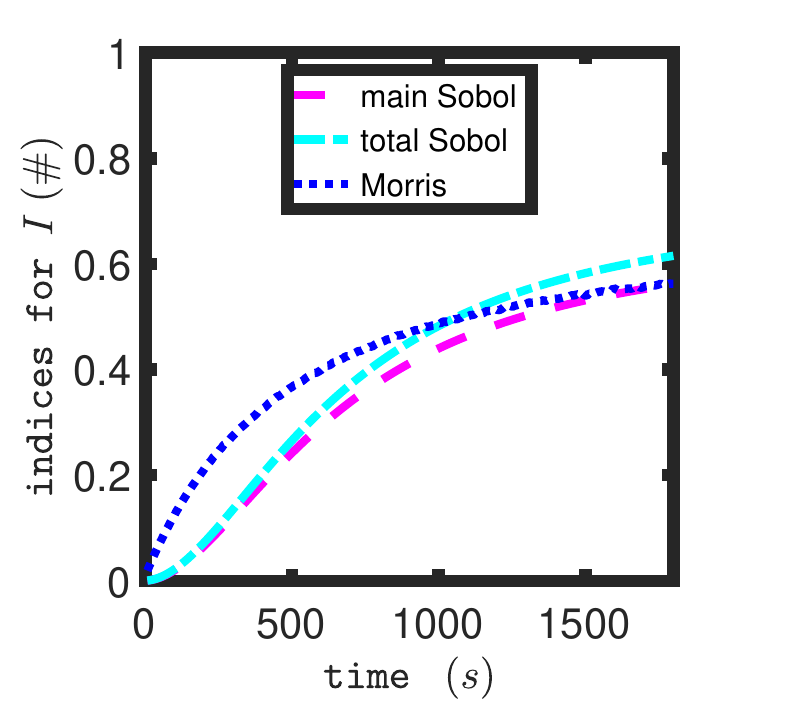}}  
  \quad
  \subfloat[$(c_\varphi)$]{\includegraphics[width=0.31\linewidth]{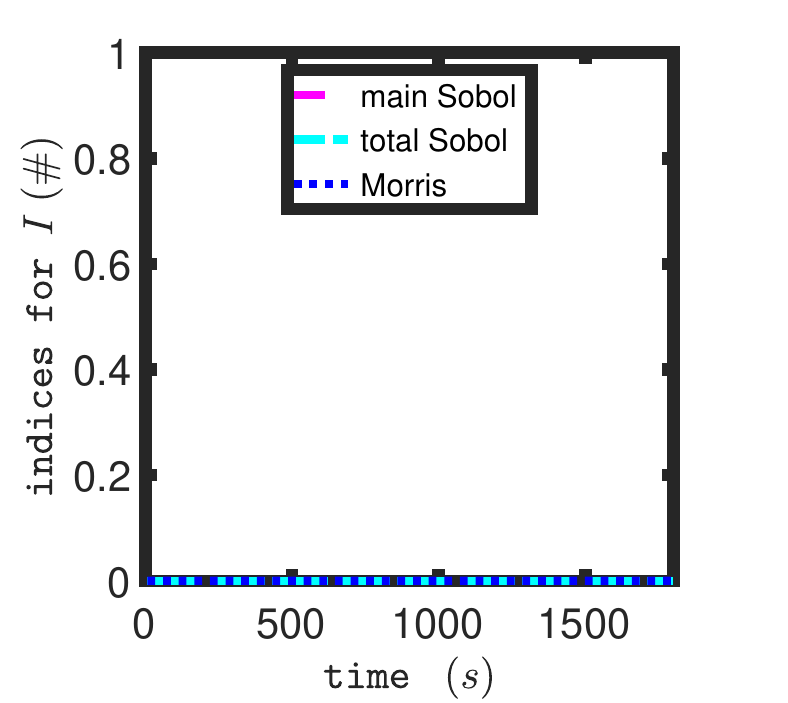}} 
  \quad
  \subfloat[$(\sigma_\varphi)$]{\includegraphics[width=0.31\linewidth]{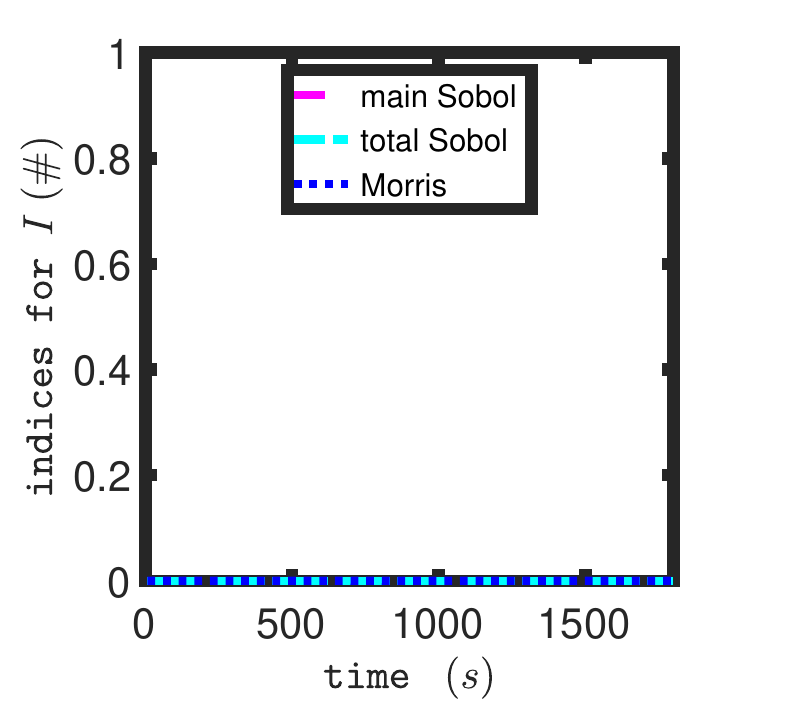}} 
   \caption{Evolution in time of the sensitivity indices of $M$.}
  \label{fig:indices-integral}
\end{figure}

We substantiate the qualitative intuition provided by the response curves via the computation of the Sobol and Morris indices. For the latter ones, we use $P=10000$ sample points and the increment $h_n = |\Gamma_n|/P$, $n=1,\dots, N$, with $|\Gamma_n|$ denoting the size of the range of the $n$-th parameter,  cf. \eqref{eq:morris-factor} and \eqref{eq:morris}.
%\sidenotelorenzo{cosa c'entra $P$ con $h_n$??}
Figure~\ref{fig:indices-median}-\ref{fig:indices-integral} display the evolution in time of the indices associated with each parameter. We preliminarily observe that, in general, Sobol and Morris indices are in good agreement. This makes the interpretation of the results easier and more reliable. In particular, it is confirmed that 
\begin{itemize}
    \item $M$ is (almost) insensitive to $\mu$ and $\sigma_\varphi$, and
    \item $I$ is insensitive to $\mu$, $\nu$, $c_\varphi$ and $\sigma_\varphi$.
\end{itemize}
All other parameters are impactful for the QoIs, with $\rho$ playing a more prominent role for later times, while the importance of the other parameters decreases.

\subsection{UQ workflow step 2: inverse UQ analysis}
\label{sec:inverse-UQ}

According to the above sensitivity analysis, the parameters $\mu$ and $\sigma_\varphi$ are irrelevant for both QoIs. Therefore, we fix them to a nominal value, namely the midpoint of their range of variability, cf. Table~\ref{tab:parametri}. It is worth mentioning also that the relevant parameters for $I$ are relevant also for $M$. Hence the Bayesian inversion can be expected to recover such parameters, leading to a substantial reduction of the uncertainty in predicting $I$, see Section~\ref{sec:forward-UQ} below. Note that we include in the calibration also $\nu$ and $c_\varphi$, that are irrelevant for $I$. Calibrating such parameters does not contribute to the uncertainty reduction of $I$, but it is necessary for the accuracy of the MAP estimate, since both $\nu$ and $c_\varphi$ are relevant for $M$, cf. Section~\ref{sec:bayes}.  

For convenience, we adapt the notation introduced in Section~\ref{sec:framework} to the new setting. The uncertain parameters are now collected into a vector of dimension $N=4$
\[
\yy = \big(\nu,\,k_{\varphi},\,\rho,\,c_{\varphi}) \in \Gamma =: \prod_{n=1}^N \Gamma_n \subset \Rset^N
\]
where $\Gamma_n := [a_n,b_n]$, $n=1,\dots, N$, are the ranges reported in Table~\ref{tab:parametri}. 

Having reduced the number of uncertain parameters, we construct new sparse-grid surrogates for the QoIs $M$ and $I$, with the same standard setup as in Section~\ref{sec:surrogates-validation}. We repeat the accuracy test and obtain similar, slightly smaller, values of the relative discrepancies (not displayed here). Therefore, we use the same sparse-grid level $w = 4$ as before. 

\begin{figure}[tp]
  \centering
  \includegraphics[width=0.42\linewidth]{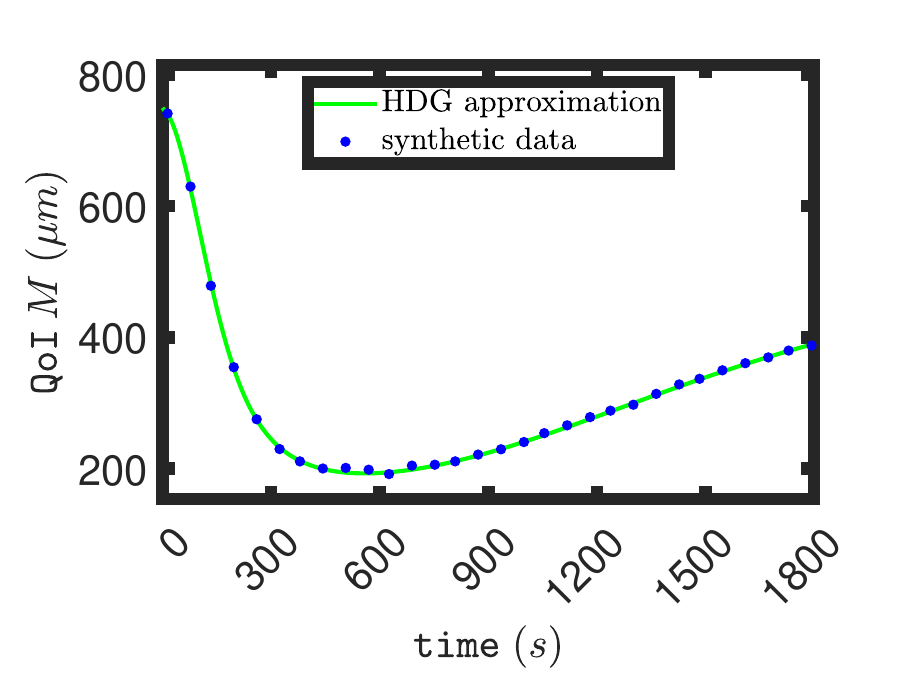}
  \qquad
  \includegraphics[width=0.4\linewidth]{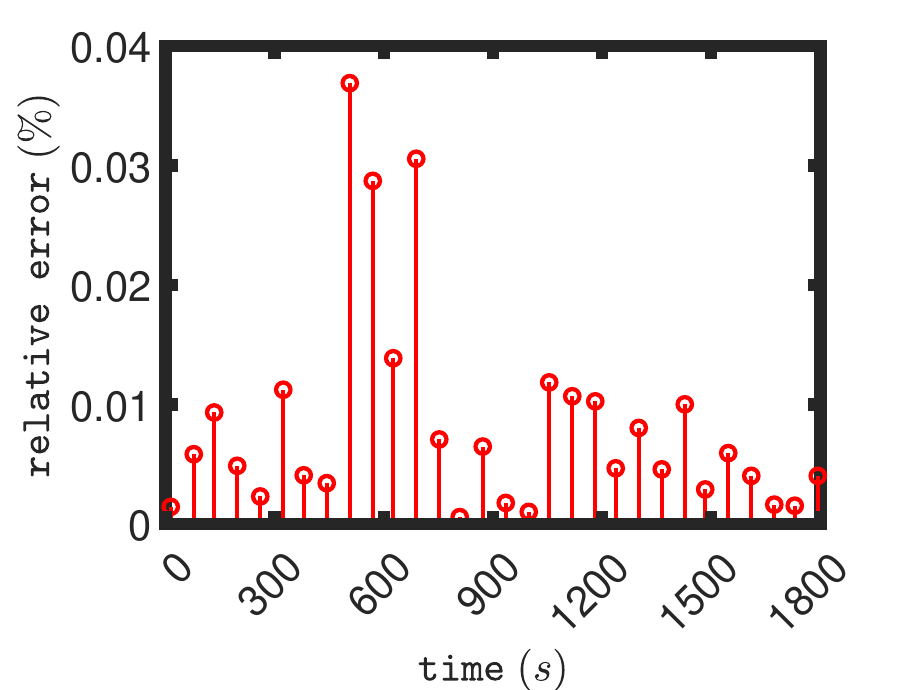}  \caption{Synthetic data for the inverse UQ analysis. Left: HDG approximation of $M$ and synthetic data. Right: relative error of the synthetic data.}
  \label{fig:synthetic-data}
\end{figure}

We generate synthetic data for the QoI $M$ via the formula in \eqref{eq:data_model}. For the reference parameter combination $\yy_{\text{true}}$, we choose the barycenter of $\Gamma$, namely
\begin{equation*}
\label{eq:y_true}
\yy_{\text{true}} = (200, 1300, 25\times 10^{-4}, 250).
\end{equation*}
We set the standard deviation of the Gaussian noise to $\sigma_\varepsilon = 2$. Data are generated for $K=30$ equispaced time instants $0 < t_1 < \dots < t_K < T$. Hereafter we report and comment on the results obtained in this setting for a single noise realization.  We observed consistent results for different $\sigma_\varepsilon$ and $K$ and/or for different noise realization. The relative error of the data used in the subsequent analysis, at the selected time instants, is displayed in Figure \ref{fig:synthetic-data}.
%\sidenotelorenzo{Pietro: dicevi che avresti salvato una collezione di risultati di backup, come sei messo su questo punto?}

We compute the MAP estimate $\yy_{\text{MAP}}$ via the minimization problem~\eqref{eq:MAP}, using the surrogates for the evaluation of $M$. We employ the derivative-free built-in Matlab routine \texttt{fminsearch} for unconstrained multivariate optimization. Since local minima are possible, we run the routine for $100$ uniformly distributed random initial points in $\Gamma$ and finally select the optimization output that realizes the minimum among all runs. The result is reported in the second column of Table~\ref{tab:statistics-parameters} below. 

\begin{figure}[tp]
  \centering
  \subfloat{\includegraphics[width=0.26\linewidth]{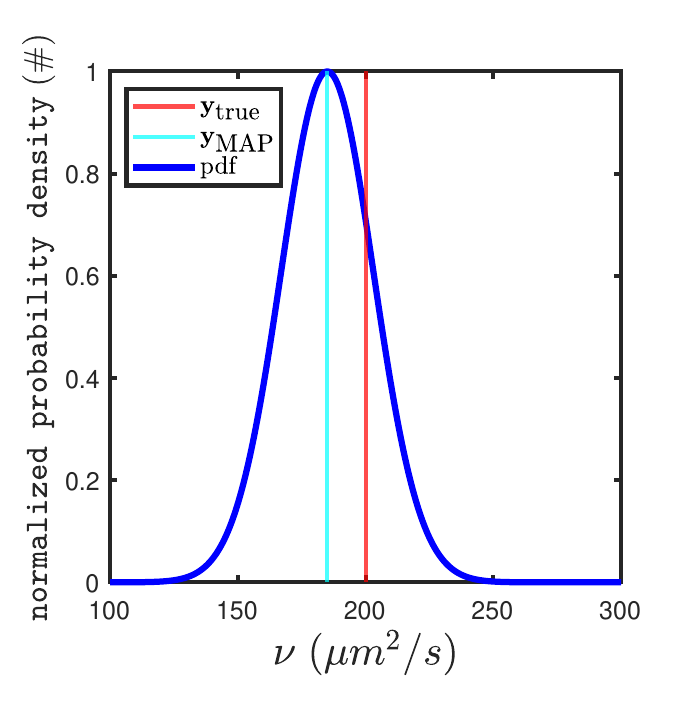}}
  \hspace{-10pt}
  \subfloat{\includegraphics[width=0.26\linewidth]{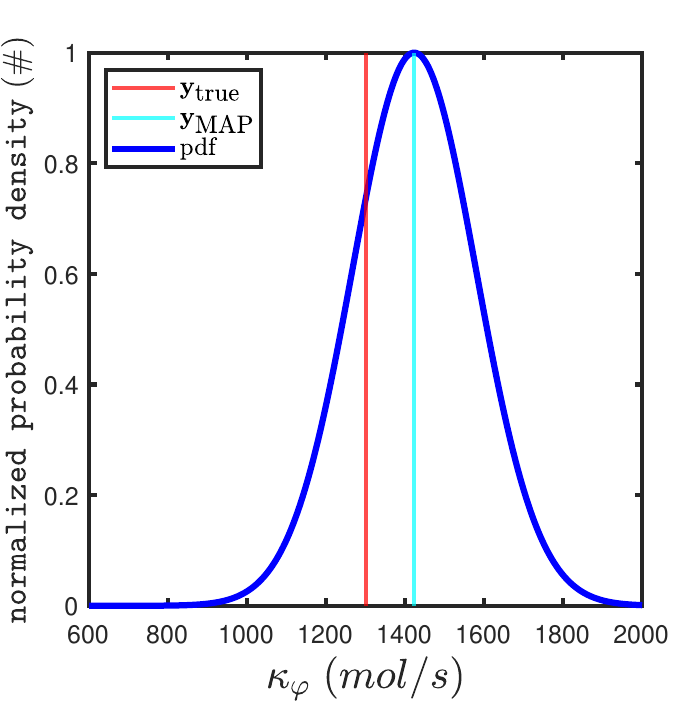}}
  \hspace{-10pt}
  \subfloat{\includegraphics[width=0.26\linewidth]{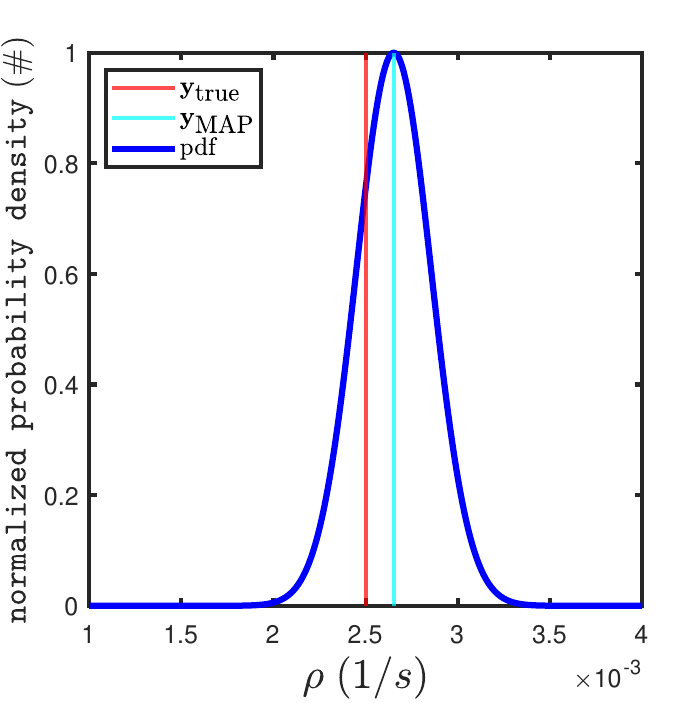}}
  \hspace{-10pt}
  \subfloat{\includegraphics[width=0.26\linewidth]{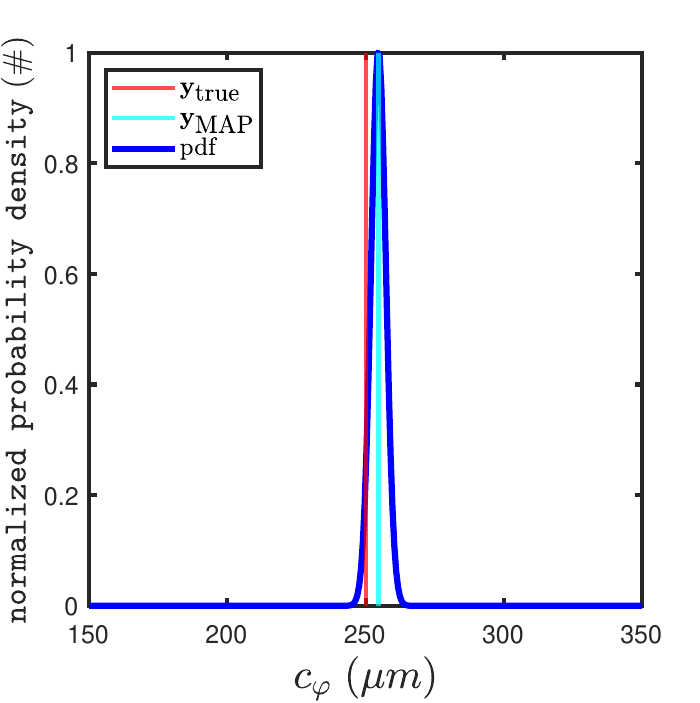}}
  \caption{MAP estimate and normalized Gaussian approximation of the marginal posterior distributions of the uncertain parameters versus the reference parameter values.
  The horizontal axis shows the prior ranges of the parameters, and therefore this figure provides a qualitative visualization of the concentration of $\rho_{\text{post}}$.}
  \label{fig:posterior-GA}
\end{figure}

With the MAP estimate at hand, we first approximate the posterior pdf $\rho_{\text{post}}$ by the multivariate Gaussian distribution with mean $\yy_{\text{MAP}}$ and the covariance matrix given by \eqref{eq:GA-covariance}. % In particular, we resort to a finite difference approximation of the Jacobian matrix in \eqref{eq:GA-Jacobian}.
The (normalized) marginal probability distributions of the uncertain parameters are reported in Figure~\ref{fig:posterior-GA}. 

Next, we verify the accuracy of the Gaussian approximation by sampling the posterior $\rho_{\text{post}}$ via the slice sampling MCMC algorithm. To this end, we first estimate the standard deviation of the noise via \eqref{eq:sigma-noise-sample-est}, using the surrogate models for the evaluation of $M$. Then, we run the built-in Matlab routine \texttt{slicesample} to generate a chain consisting of $6000$ samples distributed according to $\rho_{\text{post}}$. To enhance the quality of the samples, i.e. to increase the accuracy of the pdf estimation, it turned out to be critical selecting sufficiently high values for the burn-in and the thinning hyperparameters. Based on our tuning experience, we set the optional arguments
\[
\texttt{burnin} = 2000 \qquad \text{and} \qquad \texttt{thin} = 100
\]
in the \texttt{slicesample} routine. This entails that the algorithm produces nearly $6 \times 10^5$ samples in order to return the requested chain. Therefore, it is once more crucial approximating $M$ via the surrogate model rather than via HDG solver.fff

\begin{figure}[tp]
  \centering
  \subfloat{\includegraphics[width=0.26\linewidth]{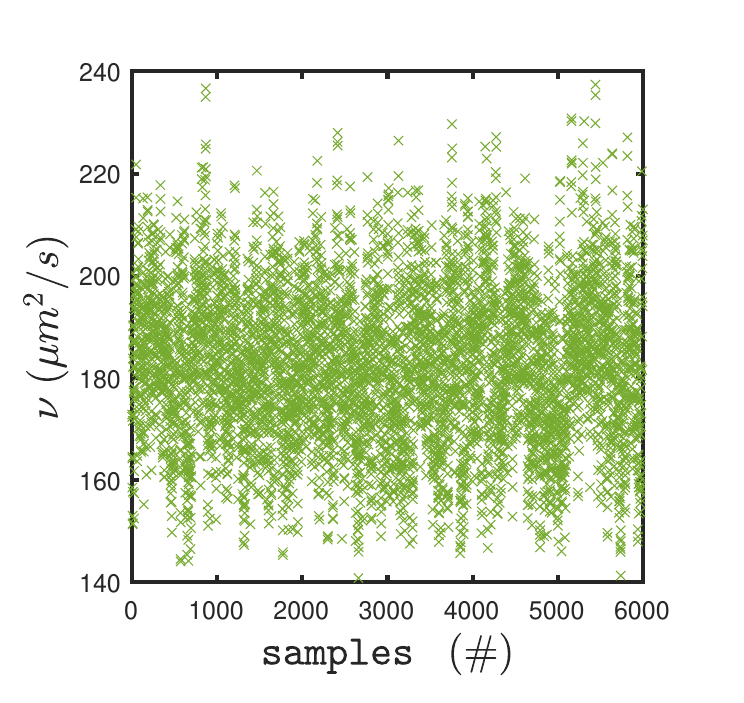}}
  \hspace{-10pt}
  \subfloat{\includegraphics[width=0.25\linewidth]{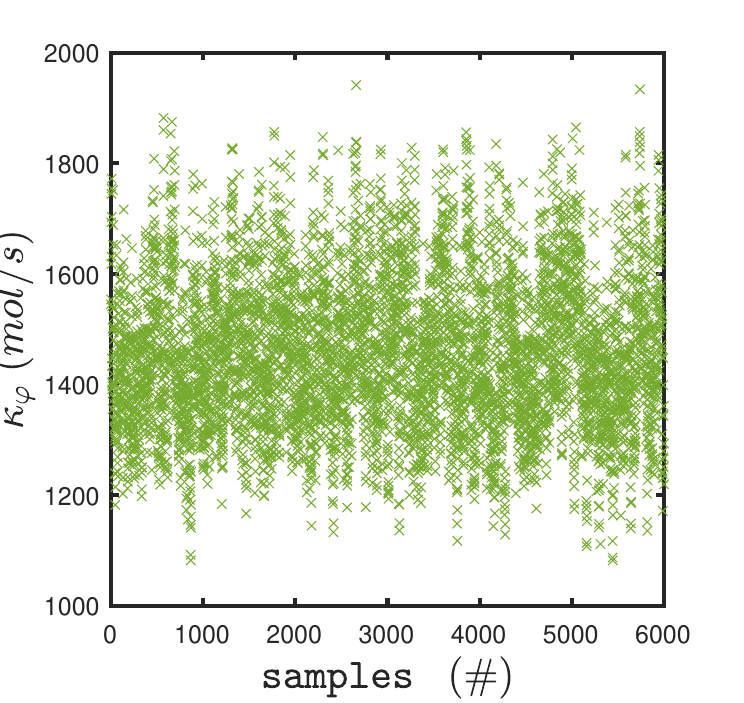}}
  \hspace{-10pt}
  \subfloat{\includegraphics[width=0.25\linewidth]{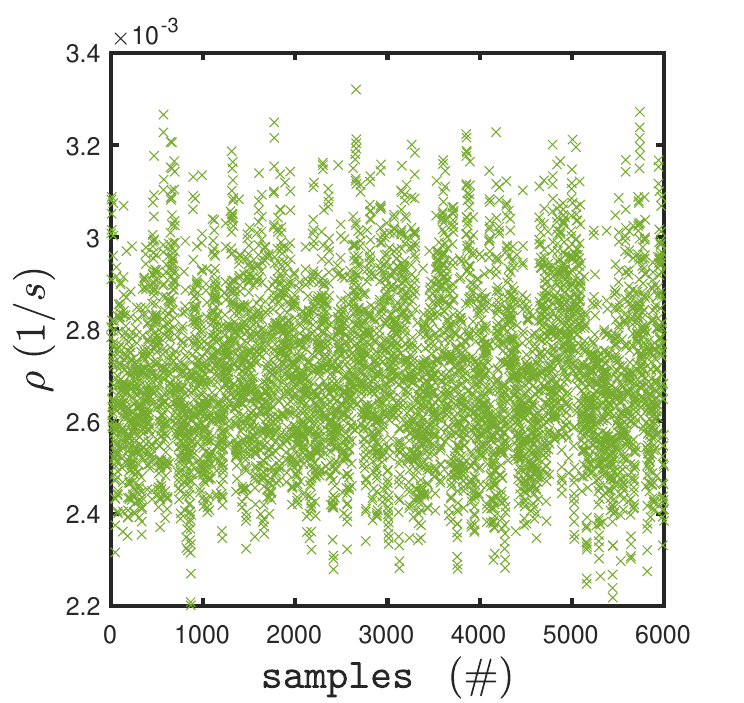}}
  \hspace{-10pt}
  \subfloat{\includegraphics[width=0.25\linewidth]{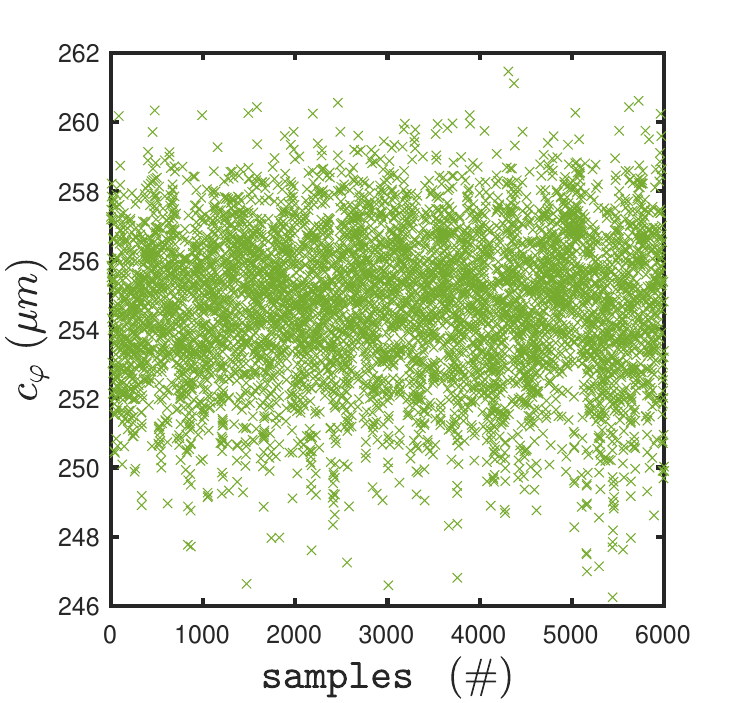}}
  \caption{Projection of the chain generated by the slicesampling MCMC algorithm on each coordinate axis in the parameter space.}
  \label{fig:MCMC-samples}
\end{figure}

\begin{figure}[tp]
  \centering
  \subfloat[$\nu$]{\includegraphics[width=0.27\linewidth]{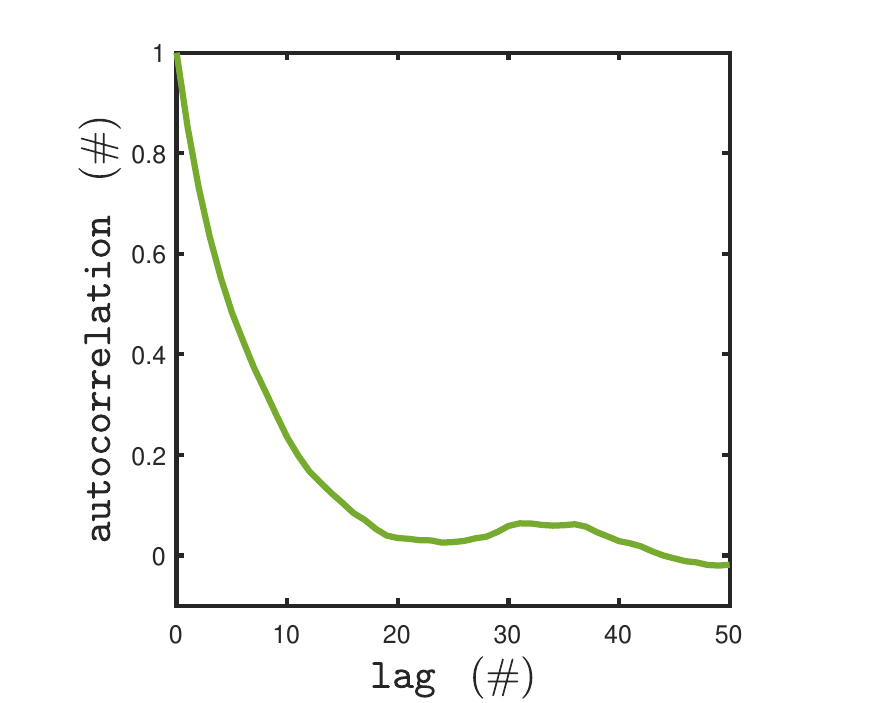}}
  \hspace{-20pt}
  \subfloat[$\kappa_\varphi$]{\includegraphics[width=0.27\linewidth]{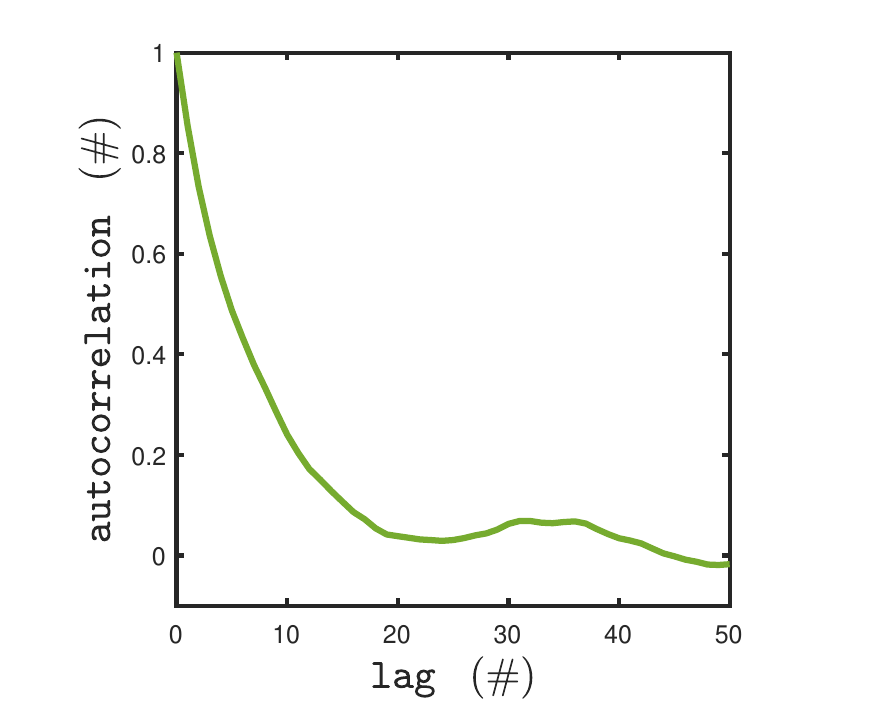}}
  \hspace{-20pt}
  \subfloat[$\rho$]{\includegraphics[width=0.27\linewidth]{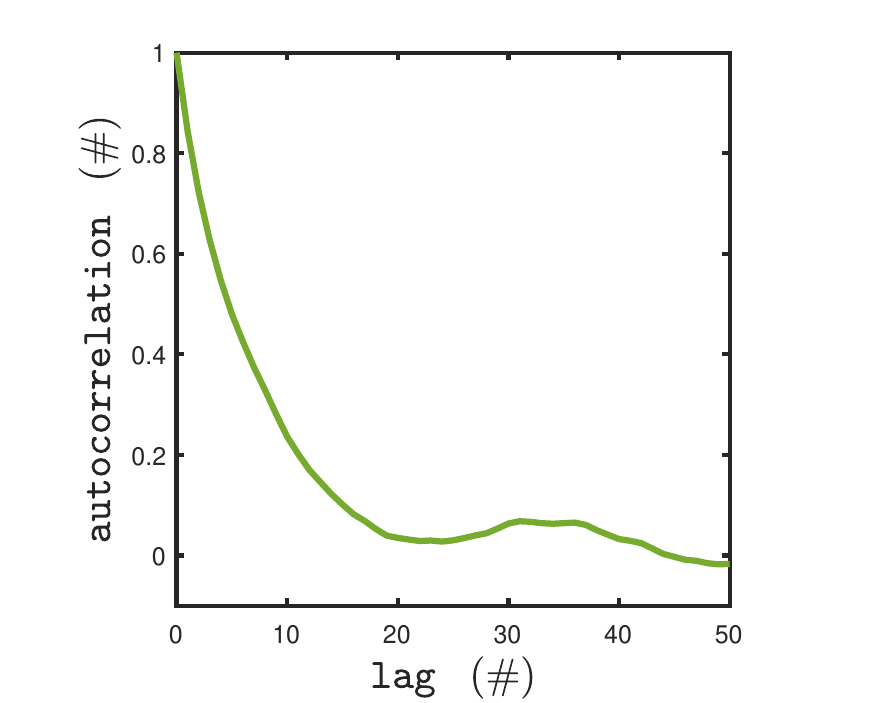}}
  \hspace{-20pt}
  \subfloat[$c_\varphi$]{\includegraphics[width=0.27\linewidth]{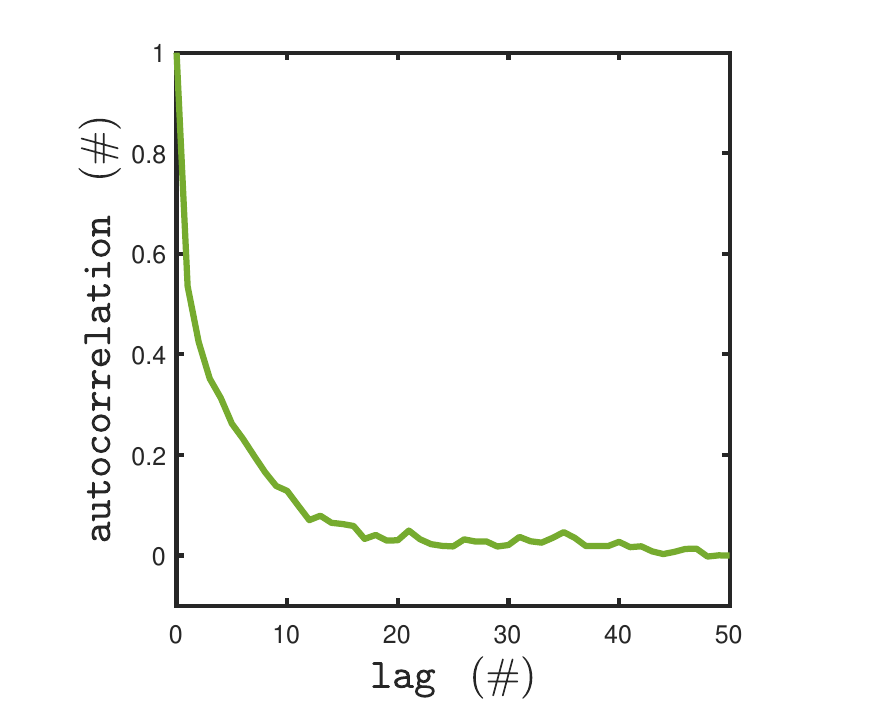}}
  \caption{Autocorrelation of the projected chains for each uncertain parameters.}
  \label{fig:MCMC-autocorrelation}
\end{figure}

We assess the quality of the chain generated by slice sampling with the aid of diagnostic plots of two types. In Figure~\ref{fig:MCMC-samples}, we plot the projection of the chain on each coordinate direction, i.e. the values taken by each parameter. In Figure~\ref{fig:MCMC-autocorrelation}, we plot the autocorrelation of each projected chain. Both figures confirm  that the chain accurately samples the posterior pdf. Indeed, the autocorrelation decays to zero relatively quickly and the samples are well-distributed in their variability range, with no visible trends. % \sidenotepietro{Non sono sicuro che `projection' sia il termine giusto in questo ambito. Nel caso, correggete}

\begin{figure}[tp]
  \centering
  \subfloat{\includegraphics[width=0.26\linewidth]{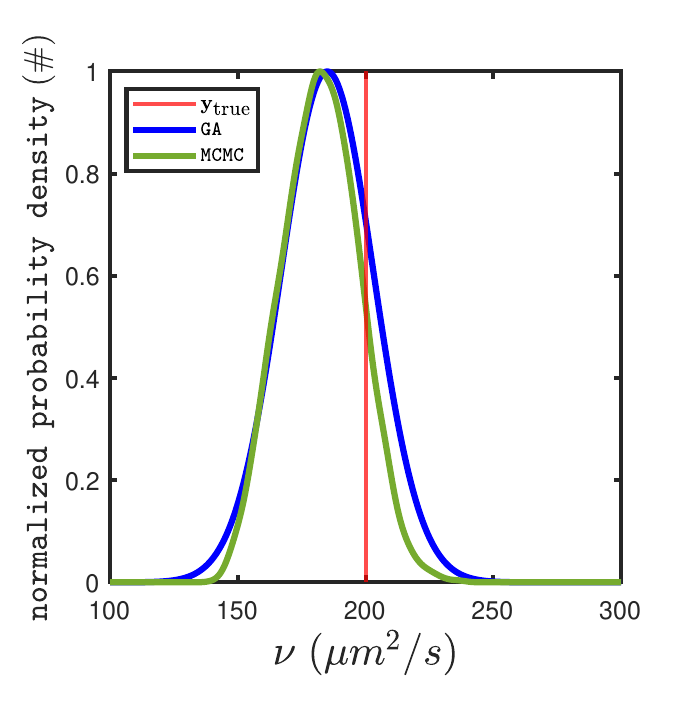}}
  \hspace{-10pt}
  \subfloat{\includegraphics[width=0.26\linewidth]{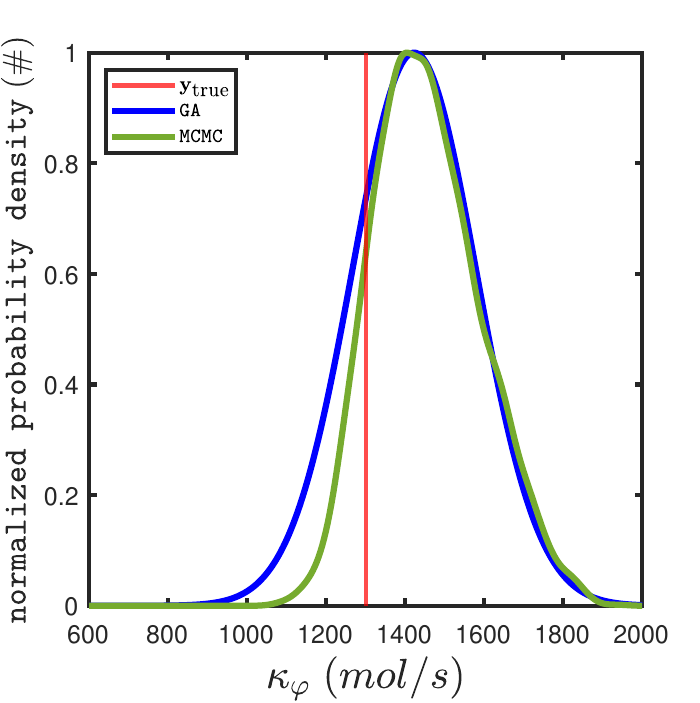}}
  \hspace{-10pt}
  \subfloat{\includegraphics[width=0.26\linewidth]{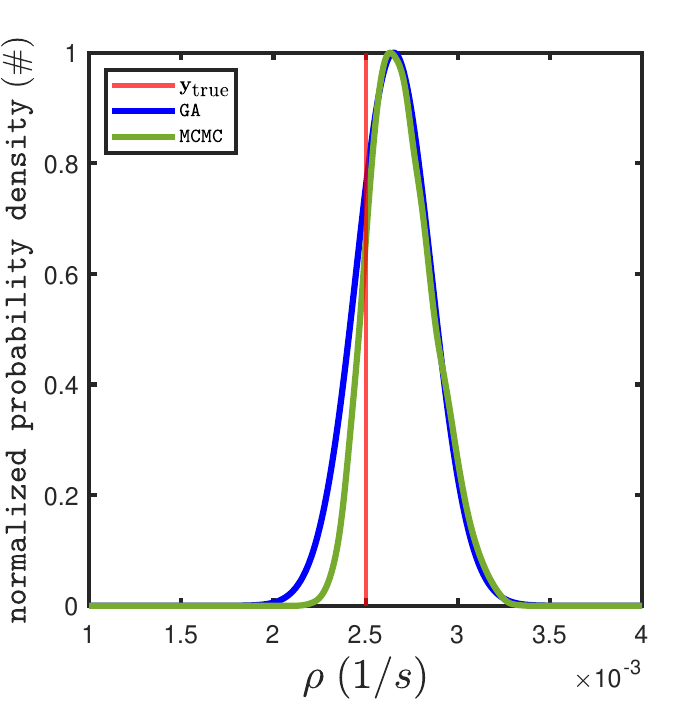}}
  \hspace{-10pt}
  \subfloat{\includegraphics[width=0.26\linewidth]{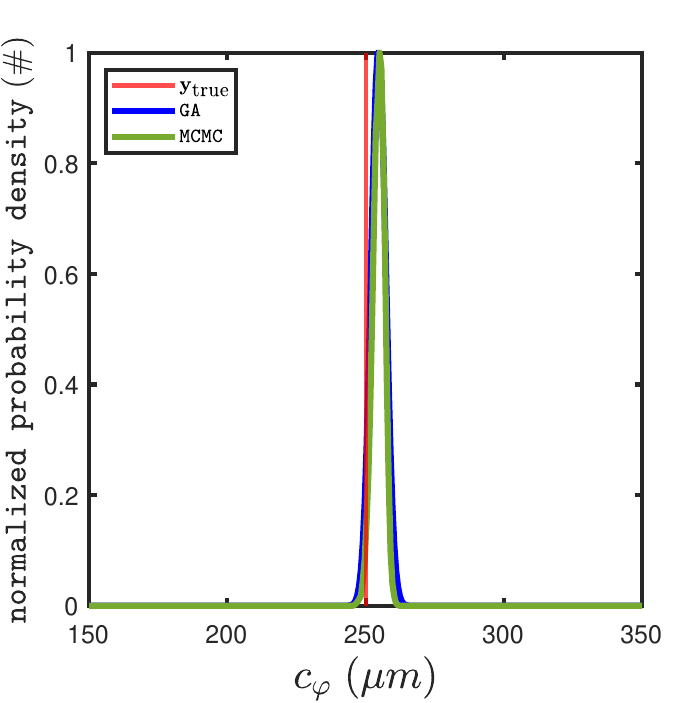}}
  \caption{Gaussian and slicesampling MCMC approximation of the marginal posterior distributions of the uncertain parameters versus the `true' parameter values.
  The horizontal axis shows the prior ranges of the parameters, and therefore this figure provides a qualitative visualization of the concentration of $\rho_{\text{post}}$.}
  \label{fig:posterior-GA-MCMC}
\end{figure}

Having verified the quality of the chain generated by the slice sampling algorithm, we compare the marginal posterior distributions with those in Figure~\ref{fig:posterior-GA}, see Figure \ref{fig:posterior-GA-MCMC}. The plots are in good agreement, with only some minor differences in the standard deviations, cf. Table~\ref{tab:statistics-parameters}. Thus, it is confirmed that the Gaussian approximation is reasonable in this case. For a better visualization of the posterior pdf, we report also the projection of the MCMC samples on all coordinate two-dimensional planes in the parameter space, see Figure~\ref{fig:MCMC-samples}. Computing the correlation matrix of the parameters via the MCMC samples yields the matrix below, which suggests an almost linear dependence among $\nu, \kappa_{\varphi}$, and $\rho$:%\sidenotepietro{Qualche commento sulle correlazioni? Sembra ci siano forti correlazioni tra i parametri. Le correlazioni calcolate con GA sono simili ma ancora più vicine a $\pm1$}
\begin{equation*}
%\label{eq:correlation-matrices}
%R_\text{GA} = \begin{pmatrix}
%1.0000 & -0.9987 & -0.9969 & -0.8317\\
%-0.9987 & 1.0000 & 0.9965 & 0.8424\\
%-0.9969 & 0.9965 & 1.0000 &0.8021\\
%-0.8317 & 0.8424 & 0.8021 & 1.0000\\
%\end{pmatrix}
R = \begin{pmatrix}
1.0000 & -0.9924 & -0.9896 & -0.6904\\
-0.9924 & 1.0000 & 0.9953 & 0.6819\\
-0.9896 & 0.9953 & 1.0000 & 0.6228\\
-0.6904 & 0.6819 & 0.6228 & 1.0000\\
\end{pmatrix}
\end{equation*}

\begin{figure}[tp]
  \centering
  \subfloat{\includegraphics[width=0.33\linewidth]{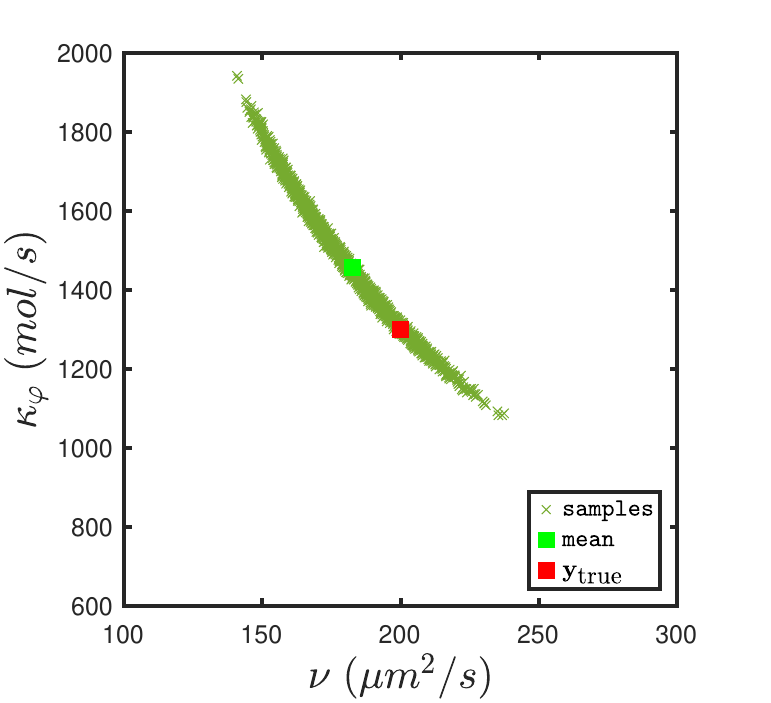}}
  \hspace{-5pt}
  \subfloat{\includegraphics[width=0.33\linewidth]{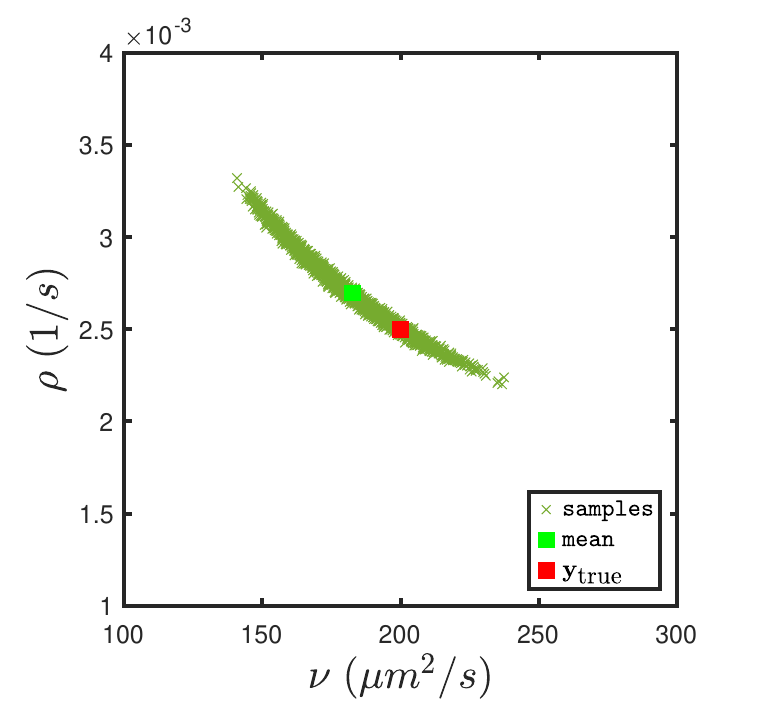}}
  \hspace{-5pt}
  \subfloat{\includegraphics[width=0.33\linewidth]{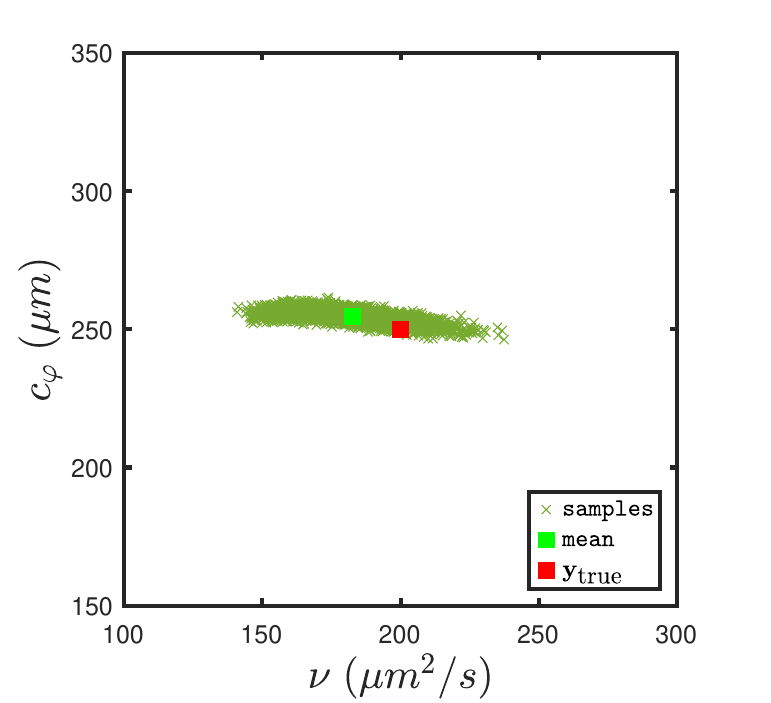}}
  \\
  \subfloat{\includegraphics[width=0.33\linewidth]{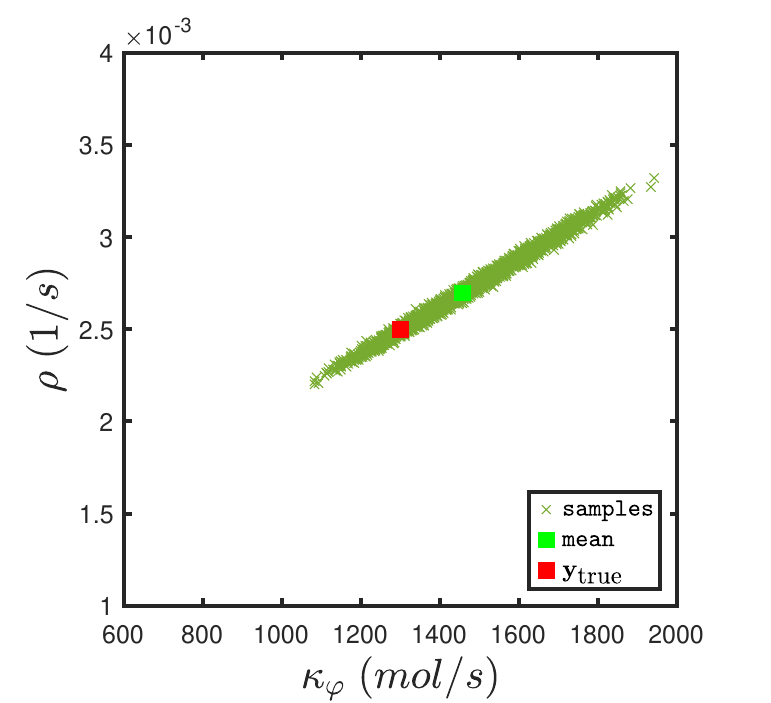}}
  \hspace{-5pt}
  \subfloat{\includegraphics[width=0.33\linewidth]{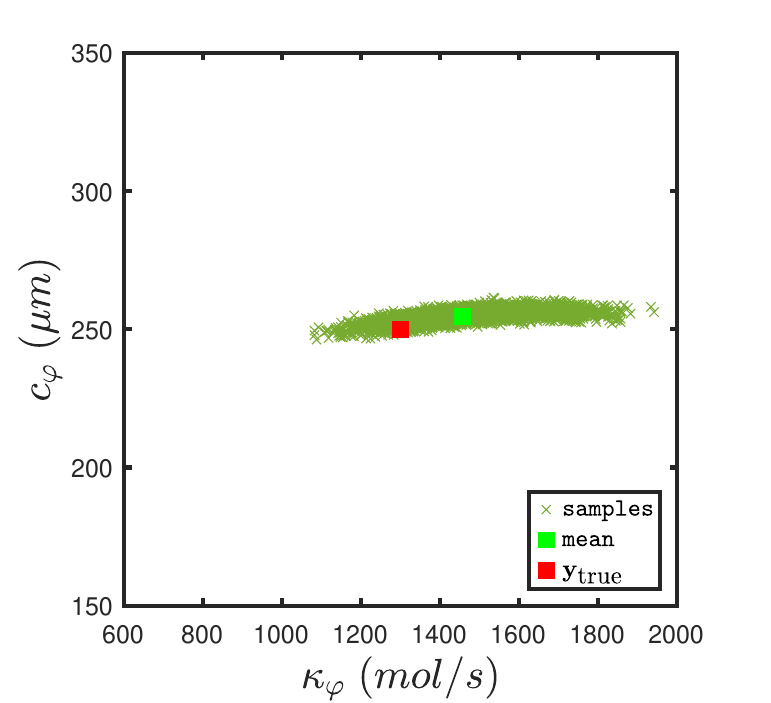}}
  \hspace{-5pt}
  \subfloat{\includegraphics[width=0.33\linewidth]{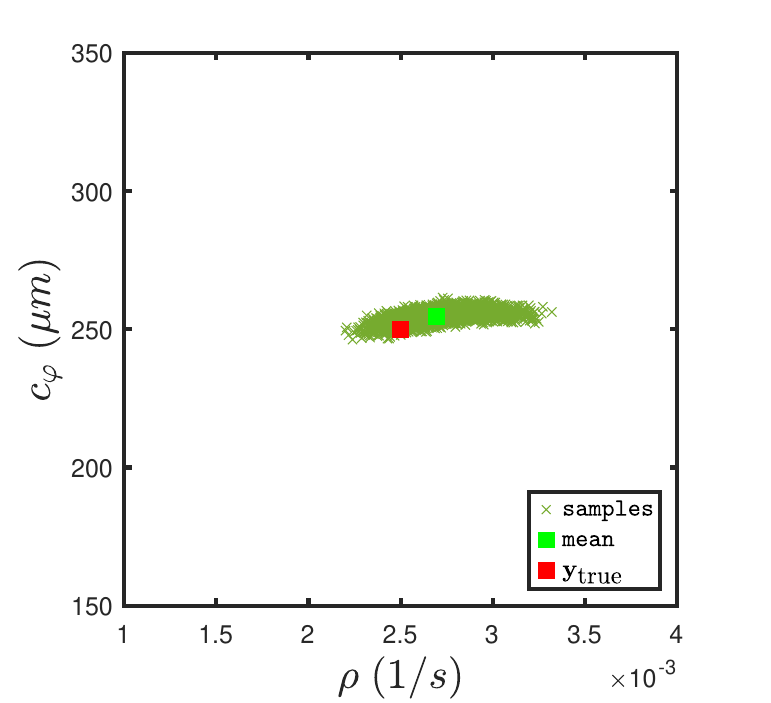}}
  \caption{Projection of the chain generated by the slicesampling MCMC algorithm on each two-dimensional coordinate plane in the parameter space.}
  \label{fig:MCMC-samples-2D}
\end{figure}

Figures~\ref{fig:posterior-GA-MCMC} and \ref{fig:MCMC-samples-2D} indicate a substantial uncertainty reduction in the distribution of the parameters. To make this observation more precise, we report some basic statistics and concentration metrics in Tables~\ref{tab:statistics-parameters} and~\ref{tab:concentration-parameters}. The latter one involves, in particular, the coefficient of variation \eqref{eq:CV} and the concentration factor \eqref{eq:CF}, so as to quantify the reduction in variability of the parameters after the Bayesian inversion step. We observe that Gaussian approximation predicts the mean (via $\yy_{\text{MAP}}$) slightly better than the slicesampling MCMC algorithm which, in turn, exhibits moderately smaller standard deviations, as visible also in Figure~\ref{fig:posterior-GA-MCMC}. This leads to moderately better concentration measures. The uncertainty reduction can be appreciated for all parameters and is more pronounced for $c_\varphi$.

\begin{table}[tp]
  \centering
  \begin{tabular}[t]{ccccccc}
    \hline 
    & \textbf{prior}  & \textbf{GA} & \textbf{MCMC} & \textbf{prior} & \textbf{GA} & \textbf{MCMC}\\
    		& \textbf{mean}  & \textbf{mean} &  \textbf{mean} & \textbf{std. dev.} & \textbf{std. dev.}  & \textbf{std. dev.}\\
    \hline
    $\nu$		&	$2.00\times 10^2$		&  $1.85 \times 10^2$ &	$1.82 \times 10^2$ &	$5.77 \times 10^1$	& $1.81 \times 10^1$ 	&	$1.50 \times 10^1$\\
    $k_\varphi$		&	$1.30 \times 10^3$		&  $1.42 \times 10^3$ &	$1.46 \times 10^3$ &	$4.04 \times 10^2$	& $1.57  \times 10^2$ 	&	$1.35 \times 10^2$\\
    $\rho$		&	$2.50 \times 10^{-3}$	&   $2.65 \times 10^{-3}$ &	$2.70 \times 10^{-3}$ &	$8.66 \times 10^{-4}$	& $2.03 \times 10^{-4}$ 	&	$1.76 \times 10^{-4}$\\
    $c_\varphi$			&	$2.50 \times 10^2$	&   $2.55 \times 10^2$ & $2.55 \times 10^2$ &	$5.77 \times 10^1$	& $2.71 \times 10^0$	&	$2.12 \times 10^0 $\\
    \hline
  \end{tabular}
  \caption{Prior and posterior statistical quantities of the uncertain parameters.}\label{tab:statistics-parameters}
\end{table}

\begin{table}[tp]
  \centering
  \begin{tabular}[t]{cccccc}
    \hline 
    & \textbf{prior} & \textbf{GA} & \textbf{MCMC} & \textbf{GA}  & \textbf{MCMC}\\
    & \textbf{CV} & \textbf{CV} & \textbf{CV} & \textbf{CF}  & \textbf{CF} \\
    \hline
    $\nu$	    & $2.89 \times 10^{-1}$	& $9.79 \times 10^{-2}$ & $8.21 \times 10^{-2}$  & $3.19 \times 10^0$ 	& $3.86 \times 10^0$	\\
    $k_\varphi$ & $3.11 \times 10^{-1}$	& $1.10 \times 10^{-1}$ & $9.28 \times 10^{-2}$ & $2.58 \times 10^0$ 	& $2.99 \times 10^0$	\\
    $\rho$		& $3.46 \times 10^{-1}$	& $7.65 \times 10^{-2}$ & $6.51 \times 10^{-2}$ & $4.27 \times 10^0$ 	& $4.93 \times 10^0$	\\
    $c_\varphi$			& $2.31 \times 10^{-1}$	& $1.07 \times 10^{-2}$ & $8.34 \times 10^{-3}$ & $2.13 \times 10^1$ 	& $2.72 \times 10^1$	\\
    \hline
  \end{tabular}
  \caption{Prior and posterior concentration measures of the uncertain parameters.}\label{tab:concentration-parameters}
\end{table}

\subsection{UQ workflow step 3: forward UQ analysis}

Once the Bayesian inversion is completed, we are in position to propagate the residual uncertainty encoded in the posterior distribution from the parameters to the QoI $I$.
For this purpose, we evaluate surrogates for the QoI $I$ on the samples the posterior pdf $\rho_{\text{post}}$, generated at the previous step
(by either of the strategies dicussed).
Although we could re-use the six-parameters surrogates for $I$ from section~\ref{sec:surrogates-validation},
it is advisable generating new ones, as we did for $M$ in Section~\ref{sec:inverse-UQ}, based only on the four relevant parameters selected after the sensitivity analysis step.
When considering the MCMC strategy, $\rho_{\text{post}}$ cannot be assumed to obey a specific distribution, therefore we use
same \textit{Clenshaw--Curtis} points used in section~\ref{sec:surrogates-validation} to build the new surrogates as well:
indeed, these points are optimal if $\rho_{\text{post}}$ were a uniform pdf, but can be used as a robust choice for a generic $\rho_{\text{post}}$.
Conversely, the (approximate) posterior distribution is explicitly known  when considering the Gaussian approximation strategy.
Therefore, we use the so-called \textit{weighted Leja} points, that are specialized for Gaussian distributions and can be expected
to deliver a more accurate approximation of $I$ in the region of the parameter space where the posterior distribution is more concentrated.

\begin{figure}[tp]
  \centering
  \subfloat[Prior]{\includegraphics[width=0.31\linewidth]{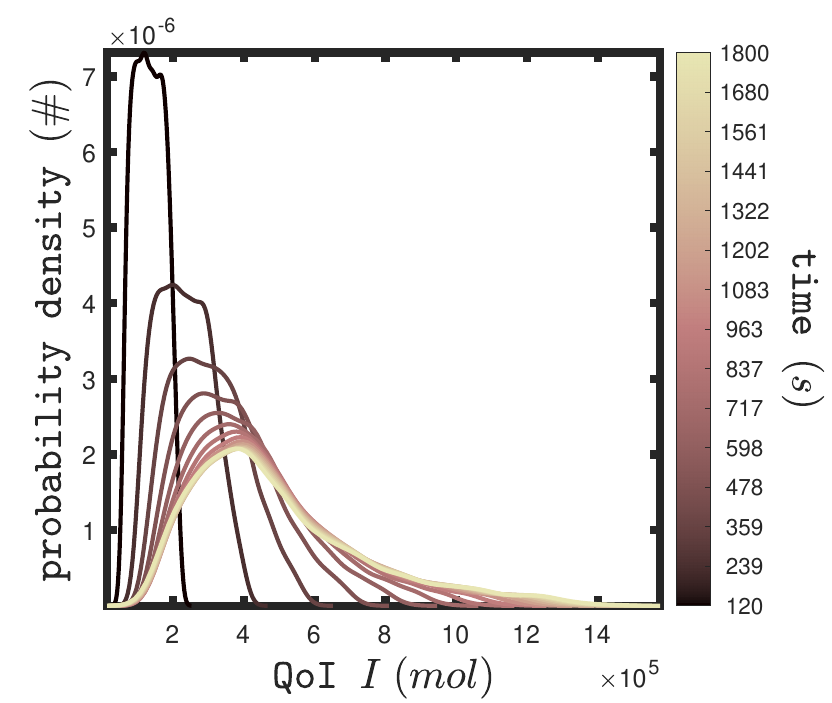}}
  \quad
  \subfloat[GA]{\includegraphics[width=0.31\linewidth]{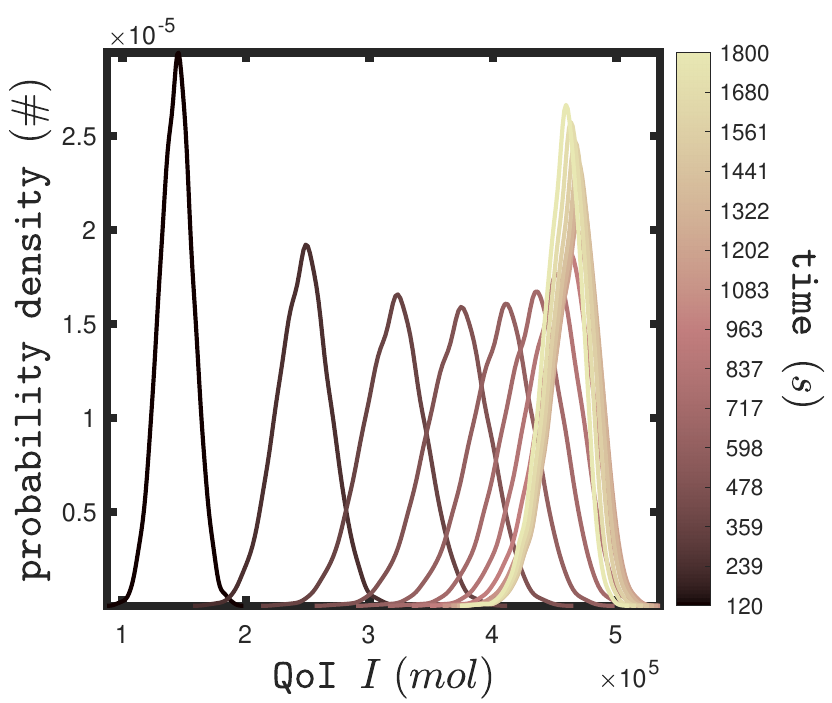}}
  \quad
  \subfloat[MCMC]{\includegraphics[width=0.31\linewidth]{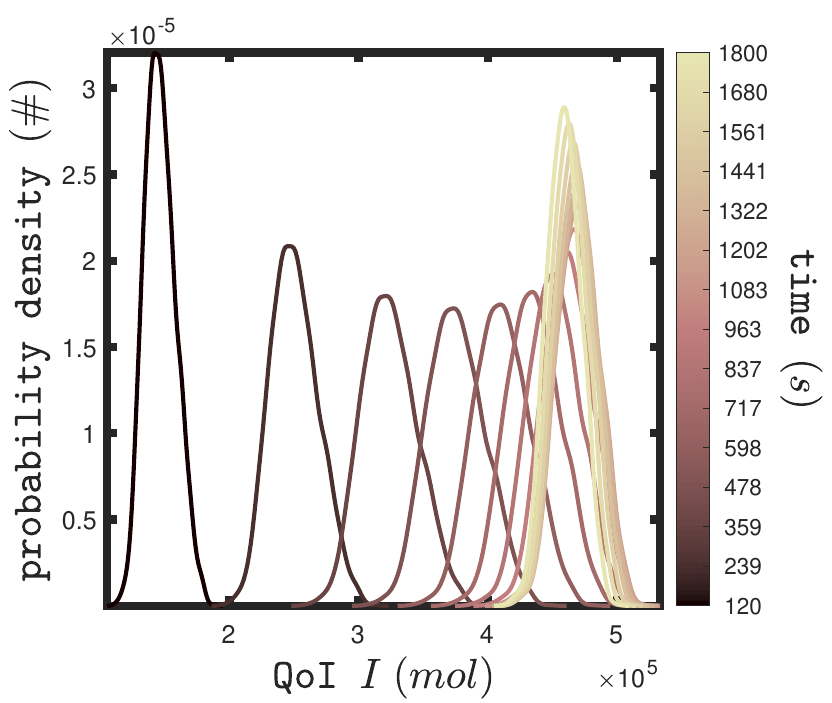}}
  \caption{Evolution in time of prior and posterior pdf of the QoI $I$.}
  \label{fig:pdf-I}
\end{figure}

\begin{figure}[tp]
  \centering
  \includegraphics[width=0.4\linewidth]{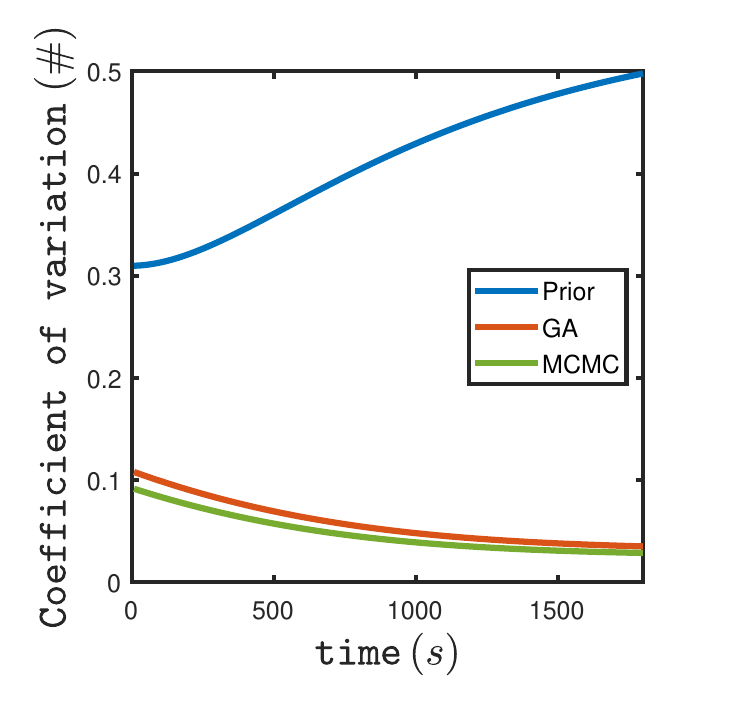}\qquad
  \includegraphics[width=0.4\linewidth]{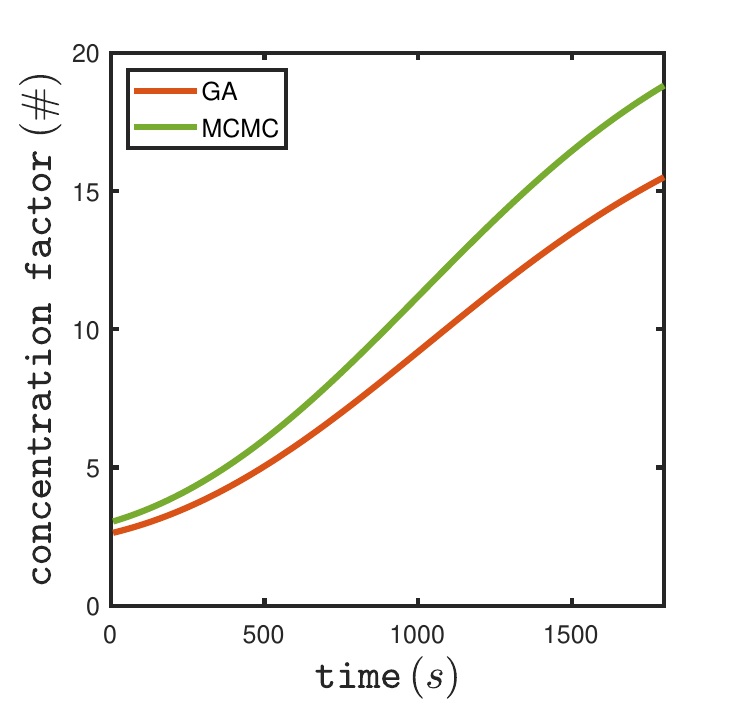}
  \caption{Evolution in time of prior and posterior concentration measures of the QoI $I$.}
  \label{fig:concentration-I}
\end{figure}

Figure~\ref{fig:pdf-I} shows the evolution in time of the prior and the posterior distribution of $I$. While there is no substantial difference between the posterior distributions obtained by Gaussian approximation and MCMC, we observe a relevant uncertainty reduction when comparing posterior and prior pdfs. (Note the different scaled on the axes.) Once again, we quantify the uncertainty reduction by means of the comparison measures CV and CF from \eqref{eq:CV} and \eqref{eq:CF}, respectively. Remarkably, the reduction is more pronounced for larger times, see Figure~\ref{fig:concentration-I}. As in Table~\ref{tab:concentration-parameters}, the approximation of the posterior distribution obtained by MCMC appears to be slightly more concentrated than by Gaussian approximation. 

\section{Conclusion and outlook}
\label{sec:conclusions}
This manuscript illustrates a complete workflow involving data-informed differential models to predict and control the dynamics of CoC experiments.
The application to a simplified one-dimensional model with synthetic data serves as a proof-of-concept, illustrating all relevant steps,
and lays the ground for the application to more realistic scenarios.
The key role played by surrogates in speeding-up the numerical simulations without substantially impacting on the overall accuracy is also demostrated.

Our future work will be devoted to the application of the proposed workflow to models able to describe CoC experiments more accurately. First of all, this calls for more sophisticated versions of the equations \eqref{eq:modello2eq} involving more unknown fields and coupling terms, as well as more realistic geometries with suitable interface conditions among the different components of devices as in Figure~\ref{fig:expGeometry}. To this end, we plan to extend the HDG solver discussed in Section~\ref{sec:HDG}, including also stabilization terms for the convection-dominated case. This is especially relevant when the uncertainty affects parameters like $\chi$ in \eqref{eq:modello2eq}. Such a solver is expected to be much more computational demanding than in this study. Therefore, we plan to resort to multi-fidelity surrogate models, such as \cite{piazzola.eal:ferry-paper}.
We will explore also some alternative to the slice sampling MCMC algorithm, such as the Hamiltonian Monte-Carlo sampling, in order to reduce the computational burden of the Bayesian inversion. As a further development of the present study, the proposed framework will be applied to the calibration of
model parameters against available experimental data from CoC experiments.

\section{Acknowledgements}
This work was supported by the Italian Ministry of Research, under the complementary actions to the NRRP “D34Health - Digital Driven Diagnostics, prognostics and therapeutics for sustainable Health care” Grant \#PNC0000001, CUP B53C22006100001.
S.B., G.B., L.T., and P.Z. are members of the Gruppo Nazionale Calcolo Scientifico-Istituto Nazionale di Alta Matematica (GNCS-INdAM).
L.T. and P.Z. have been partially supported by the PRIN 2022 PNRR project ``Uncertainty Quantification of coupled models for water flow and contaminant
transport'' (No. P2022LXLYY), financed by the European Union - Next Generation EU.

%\sidenotepietro{Bibliografia controllata, ripulita e uniformata. Se volete aggiungere qualcosa, mantenente lo stile del file}

\bibliographystyle{unsrt}
\bibliography{fontibib}

\end{document}